\documentclass[proc]{edpsmath}
\usepackage{hyperref}
\usepackage[numbers]{natbib}
\usepackage{float}
\usepackage{graphicx}
\usepackage{subfigure}
\usepackage{xspace}
\usepackage{algorithmic, algorithm}

\newcommand{\R}{\mathbb{R}}
\newcommand{\X}{\mathbb{X}}
\newcommand{\Y}{\mathbb{Y}}

\newcommand{\p}{\mathbb{P}}
\newcommand{\N}{\mathbb{N}}
\newcommand{\mN}{\mathcal{N}}
\newcommand{\mM}{\mathcal{M}}
\newcommand{\mZ}{\mathcal{Z}}
\newcommand{\mO}{\mathcal{O}}

\begin{document}

\title{Sequential Bayesian inference for implicit hidden Markov models and current limitations} 

\author{Pierre E. Jacob}\address{Department of Statistics, University of Oxford, pierre.jacob@stats.ox.ac.uk.  }
\dedicated{\it Dedicated to Philip Perry} 

\begin{abstract} 
    Hidden Markov models can describe time series arising in various fields of science, by treating the data as noisy 
    measurements of an arbitrarily complex Markov process. Sequential Monte Carlo (SMC) methods have become standard
    tools to estimate the hidden Markov process given the observations and a fixed parameter value. We review some of the recent
    developments allowing the inclusion of parameter uncertainty as well as model uncertainty. The shortcomings
    of the currently available methodology are emphasised from an algorithmic complexity perspective.  
    The statistical objects of interest for time series analysis are illustrated on a toy ``Lotka-Volterra'' model used in population ecology. Some open challenges are discussed regarding the scalability of the reviewed methodology to longer time series, higher-dimensional state spaces and more flexible models.
\end{abstract}

\begin{resume} 
    Les mod\`eles \`a cha\^ine de Markov cach\'ee permettent de d\'ecrire les s\'eries temporelles de divers domaines scientifiques, en traitant les donn\'ees comme
    des mesures bruit\'ees d'un processus de Markov arbitrairement complexe. Les m\'ethodes de Monte Carlo s\'equentielles (SMC) sont devenues des outils standards
    pour l'estimation du processus de Markov cach\'e sachant les observations et une valeur fix\'ee du param\`etre. Nous passons en revue quelques unes des 
    r\'ecentes avanc\'ees permettant de prendre en compte l'incertitude sur le param\`etre ainsi que l'incertitude sur le mod\`ele. Les limites de 
    la m\'ethodologie actuelle sont discut\'ees sous l'angle de la complexit\'e algorithmique.
    Les objets statistiques d'int\'er\^et pour l'analyse des s\'eries temporelles sont illustr\'es sur un mod\`ele jouet de type ``Lotka-Volterra'' utilis\'e
    en \'ecologie des populations. Quelques questions ouvertes sont finalement pos\'ees concernant
    l'extension de la m\'ethodologie pr\'esent\'ee pour traiter des s\'eries de donn\'ees plus longues, des espaces d'\'etats de dimension plus grande et des mod\`eles plus flexibles.
\end{resume}

\maketitle

%%-----------------------------
%%      your text
%%-----------------------------
\section{Setting\label{sec:setting}}

\subsection{Hidden Markov models \label{sec:sub:hmm}}

Hidden Markov models constitute a very flexible class of models for time
series data. Consider time series made of real-valued vectors
$y_t\in\Y\subset\R^{d_y}$ for a countable collection of times $t\in\{t_1,
\ldots, t_n, \ldots\}$ and some integer $d_y$. For simplicity we consider integer-valued times.
Hidden Markov models propose to treat the observations
$(y_t)_{t\in\N}$ as if they were arising from noisy measurements of a latent
Markov chain $(x_t)_{t\in\N}$.  First, the distribution of a Markov chain
$(x_t)_{t\in\N}$ living in $\X\subset\R^{d_x}$, for some integer $d_x$, is specified
through the distribution $\mu(dx_0)$ of its initial state and the conditional
distribution of each successive state $f(dx_{t}\mid x_{t-1})$, which is called
the transition distribution. Then the model specifies the distribution
$g(dy_t\mid x_t)$ of each observation given the current hidden state, which is
called the measurement distribution.  Finally all those distributions are
parametrized by a vector $\theta$ living in a set $\Theta\subset
\R^{d_\theta}$ for some integer $d_\theta$. We explicitly write the
parameter $\theta$ in $\mu(dx_0 \mid \theta)$, $f(dx_t \mid x_{t-1},\theta)$
and $g(dy_t \mid x_t,\theta)$, and a model $\mM$ refers to the collection of
objects $\{\Theta, \mu, f, g\}$. 
In the following, for a sequence $(v_t)_{t\in\mathbb{N}}$ (resp. $(v^t)_{t\in\mathbb{N}}$) and $0\leq s\leq t$,
the vector $(v_s,\ldots,v_t)$ (resp. $(v^s,\ldots,v^t)$) is denoted by $v_{s:t}$ (resp. $v^{s:t}$).

\begin{figure}
    \centering
    \includegraphics[width=0.5\textwidth]{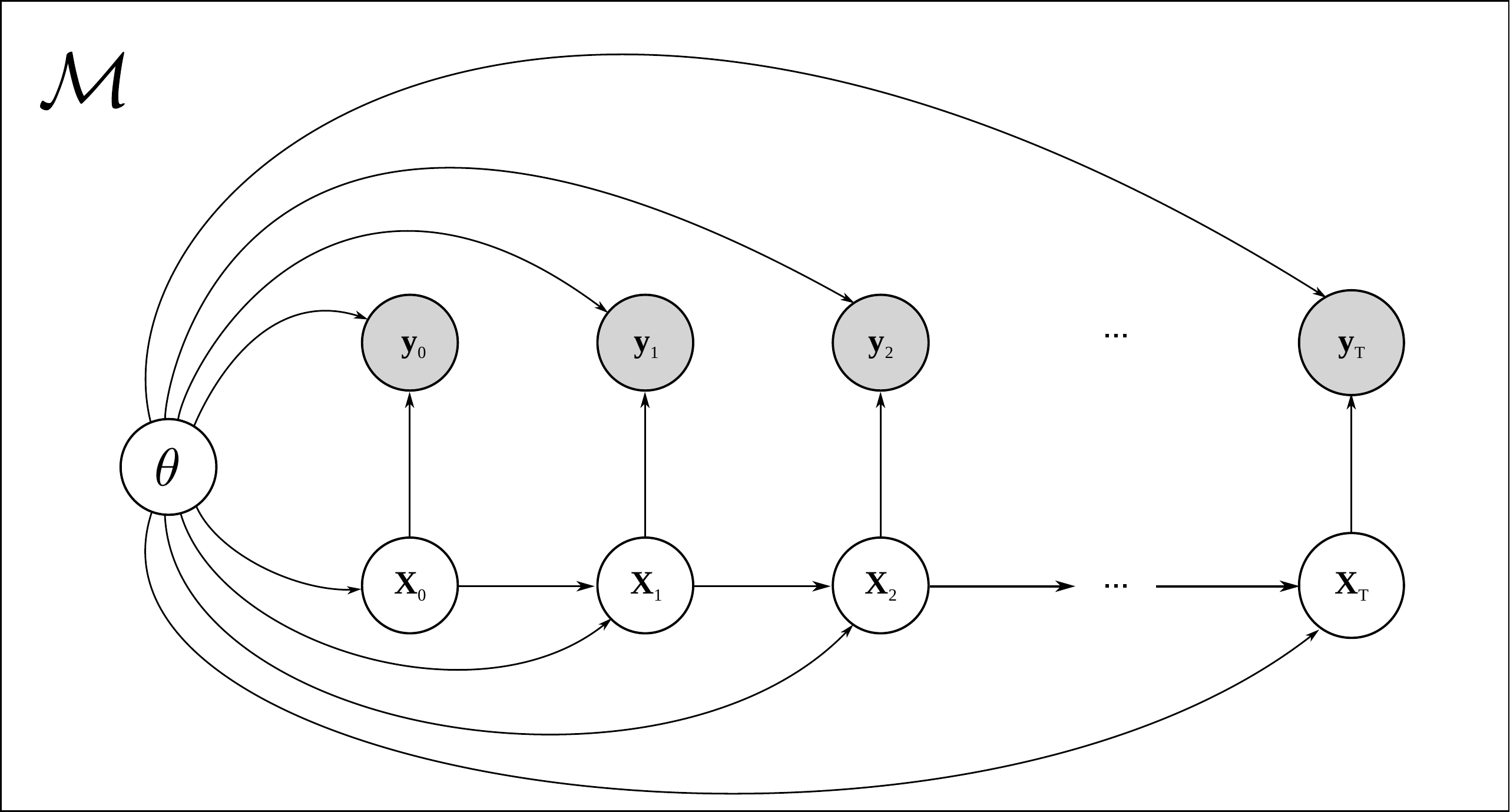}
    \caption{Graphical representation of the variables defined by a hidden Markov model $\mM$.}
    \label{fig:hmm}
\end{figure}

\begin{figure}
    \centering
    \includegraphics[width=0.6\textwidth]{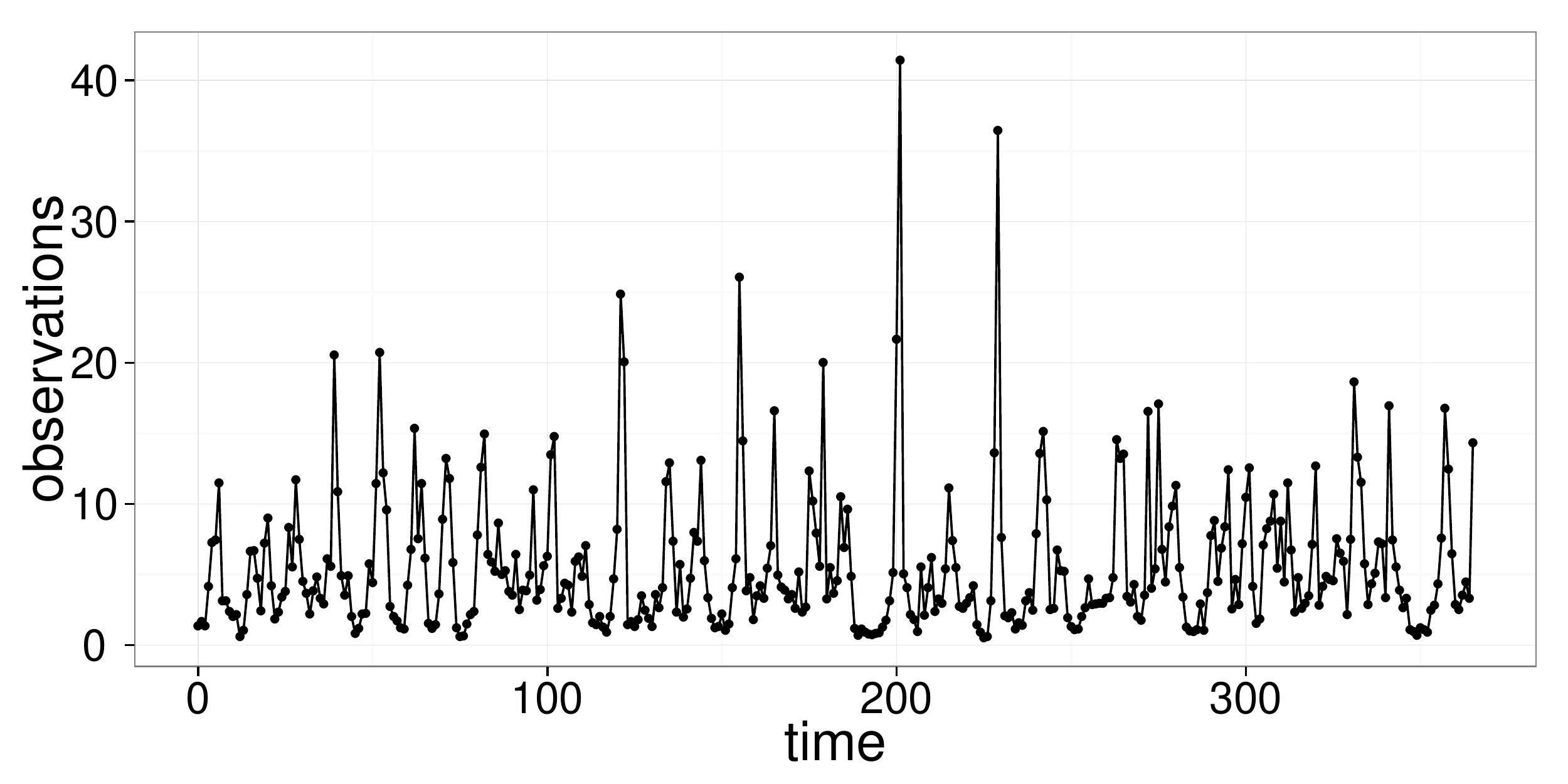}
    \caption{A time series of $365$ observations generated according to the phytoplankton--zooplankton model described in Section~\ref{sec:sub:implicit}.
             The observations represent daily measurements of phytoplankton concentrations in a volume of water.}
    \label{fig:pz:obs}
\end{figure}

Hidden Markov models are often represented by dependency graphs such as Figure
\ref{fig:hmm}, although $\theta$ is typically omitted.  The graph
indicates that 
the law of $y_t$ given $x_t$ and $\theta$ is independent of the other variables,
the law of $x_t$ given $x_{t-1}$, $y_t$, $x_{t+1}$ and $\theta$ is independent of the other variables,
and so on.  Figure~\ref{fig:pz:obs} shows a time series
generated by a hidden Markov model to be specified in Section~\ref{sec:sub:implicit}.  

After having obtained some data, the user comes up with one or several candidate
models, denoted by $\mM^{(1)}, \ldots, \mM^{(M)}$, for some integer $M$.  Hidden Markov models are
routinely applied to the modeling of volatility of financial assets, of
meteorological time series such as wind speed and direction, or of population
growth in ecology (see \cite{CapMouRyd,DouFreiGor} for a variety of applications). The
design of a model preferably takes into account as much knowledge as possible
on the phenomenon under study, and a particular model developed for one
phenomenon is unlikely to be of use for others. There are exceptions: for instance,
stochastic volatility models have been applied to pollution data in
\cite{achcar2011using}. 

In the rest of the article we assume that a
collection of models is given to us.  Then the following questions naturally
arise. 
\begin{itemize}
    \item For each model, how do the data inform about the parameters?
    \item For each model, how do the data inform about the latent Markov process?
    \item How do the data inform about the choice of a model? 
    \item How to predict future observations?
\end{itemize}
If one can simulate synthetic datasets from a model given any parameter value, then by trial and error,
one can gather some intuition on how models perform.
Intuitively, a model that generates synthetic time series that resemble actual data, at least for some parameter value,
is believed to be a good model, for instance in the sense that one would hope its prediction of future observations to be reliable.

Statistical inference is concerned with formalizing and
automatizing this ad hoc procedure, by making a principled connection between data
and models.  In Section~\ref{sec:sub:objects} the Bayesian framework is shown to transform
the above questions into various integrals. In Section~\ref{sec:sub:implicit} we introduce the notion of implicit model,
on which we focus thereafter. 
In Section~\ref{sec:sub:blackbox} we state some desired properties of statistical computing methods for time series models.  Section~\ref{sec:methods}
reviews some numerical methods compatible with implicit models, called ``plug and play'' methods,
and discusses whether they meet the desired requirements. 
In particular, sequential Monte Carlo (SMC) and particle Markov chain Monte Carlo methods are introduced.
In Section~\ref{sec:smcsamplers}
we describe a recently proposed method called SMC$^2$  to perform Bayesian inference
sequentially and mention its shortcomings.  Section~\ref{sec:numerics}
illustrates the method on a population growth model. Finally Section~\ref{sec:discussion}
discusses the reviewed methodology and some open challenges.

\subsection{Objects of inference\label{sec:sub:objects}}

To begin with, the model $\mM$ and a particular parameter value $\theta$ are
considered fixed, and the goal is to estimate the distribution of the hidden
process given observations. Filtering refers to the task of
estimating at time $t$ the current state $x_t$ given the available observations $y_{0:t}$.
The filtering distribution $p(dx_t\mid y_{0:t}, \theta)$ is obtained using Bayes rule as
\begin{equation}
    \int_\X \varphi(x_t) \; p(dx_t\mid y_{0:t}, \theta) = \frac{1}{p(y_{0:t}\mid\theta)} \int_{\X^{t+1}} \varphi(x_t) \underbrace{\left(\mu(dx_0\mid \theta) \prod_{s=1}^t f(dx_s\mid x_{s-1},\theta)\right)}_{\text{law of the latent Markov chain}} 
\underbrace{\left(\prod_{s=0}^t g(y_s\mid x_s,\theta)\right)}_{\text{conditional law of the observations}} 
    \label{eq:filtering}
\end{equation}
for any test function $\varphi$. In the rest of the article, $\varphi$ refers to a generic test function, defined either on $\X$, $\Y$ or $\Theta$. The normalizing constant $p(y_{0:t}\mid \theta)$ in 
Eq.~\eqref{eq:filtering} is called the marginal likelihood of $\theta$ and we come back to it in Eq.~\eqref{eq:likelihood}. Filtering sometimes refers to the distribution
$p(dx_{0:t}\mid y_{0:t}, \theta)$ of the path (or ``trajectory'') $x_{0:t}$ given $y_{0:t}$.
%****** 
Prediction refers to the inference of both a future observation $y_{t+k}$ and a future state $x_{t+k}$ given the current observations $y_{0:t}$, for some $k\geq 1$.
Whether the interest lies on future states or future observations depends on the application.
Using the Markov property, prediction is obtained as a by-product of the filtering distribution.
Denoting the state predictive distribution by $p(dx_{t+k} \mid y_{0:t}, \theta)$, it
can be written as 
\begin{equation}
\int_\X \varphi(x_{t+k}) \; p(dx_{t+k} \mid y_{0:t}, \theta) = 
\int_{\X^{k+1}} \varphi(x_{t+k}) \underbrace{p(dx_t\mid y_{0:t}, \theta)}_{\text{filtering law}} \underbrace{\left( \prod_{s=t+1}^{t+k} f(dx_s\mid x_{s-1},\theta)\right)}_{\text{conditional law of the future Markov chain}} .
    \label{eq:pred:states}
\end{equation}
The predictive distribution of $y_{t+k}$ given $y_{0:t}$, denoted by $p(dy_{t+k}\mid y_{0:t}, \theta)$, is then derived as
\begin{equation}
\int_\Y \varphi(y_{t+k}) \; p(dy_{t+k}\mid y_{0:t}, \theta) = 
\int_\Y \int_\X \varphi(y_{t+k}) \; \underbrace{p(dx_{t+k}\mid y_{0:t}, \theta)}_{\text{state prediction}} \; \underbrace{g(dy_{t+k}\mid x_{t+k},\theta)}_{\text{measurement}}. 
    \label{eq:pred:obs}
\end{equation}
Finally smoothing refers to inference of a past state $x_{t-k}$ given $y_{0:t}$, for some $1\leq k \leq t$.

In many realistic situations one does not know which parameter value $\theta$ to plug in the model, even if the interest lies
in filtering the hidden process; in other settings parameters are themselves the objects of interest.
To learn about the parameters from the observations, a ``prior'' probability distribution $\pi_\theta$ is given to the parameter
$\theta$ and the goal is to estimate the ``posterior'' distribution given the observations.  
The marginal likelihood of $\theta$ is the probability density function of the distribution of the data given
$\theta$,
evaluated at the observations $y_{0:t}$:
\begin{equation}
p(y_{0:t}\mid\theta) = \int_{\X^{t+1}} {\mu(dx_0\mid \theta) \prod_{s=1}^t f(dx_s\mid x_{s-1},\theta)} {\prod_{s=0}^t g(y_s\mid x_s,\theta)}.
    \label{eq:likelihood}
\end{equation}
The marginal likelihood of $\theta$ was the normalizing constant in Eq.~\eqref{eq:filtering};
the ``full likelihood'' sometimes refers to the joint probability density function of $x_{0:t}$ and $y_{0:t}$ given $\theta$.
Then the posterior distribution $\pi_{\theta,t}$ of $\theta$ given $y_{0:t}$
is defined by Bayes rule as 
\begin{equation}
\int_\Theta \varphi(\theta) \pi_{\theta,t}(d\theta) = \frac{1}{\int_\Theta p(y_{0:t}\mid \theta)\pi_\theta(d\theta)} \int_\Theta \varphi(\theta) p(y_{0:t}\mid \theta) \pi_\theta(d\theta).
    \label{eq:posterior}
\end{equation}
The normalizing constant in Eq.~\eqref{eq:posterior} will be useful for model comparison, 
and we come back to it in Eq.~\eqref{eq:evidence}.
By taking parameter uncertainty into account, one can redefine the tasks of filtering, prediction and smoothing.
For instance, filtering under parameter uncertainty refers to the distribution $p(dx_t\mid y_{0:t})$ of the
current hidden state $x_t$ given $y_{0:t}$, averaged over all possible parameters:
\begin{equation}
    \int_\X \varphi(x_t) \; p(dx_t\mid y_{0:t}) = \int_\Theta \int_{\X} \varphi(x_t)\underbrace{p(dx_t\mid y_{0:t}, \theta)}_{\text{filtering given parameter}} 
    \underbrace{\pi_{\theta,t}(d\theta)}_{\text{posterior on parameter}} .
    \label{eq:filtering:uncertainty}
\end{equation}
Likewise one can be interested in filtering over the full paths, prediction and smoothing under parameter uncertainty.

Once parameter uncertainty is taken into account, the next source of uncertainty is at the model level.
When several models $\mM^{(1)}, \ldots, \mM^{(M)}$ are available, there are various ways to use the observations to compare models (see Chapter 7 of \cite{Robert:Bayesian:book},
Chapter 6 of \cite{bernardo2009bayesian}).
A building block of model comparison is the ``model evidence''. The evidence of $\mM^{(m)}$ is defined as the
normalizing constant $\mZ^{(m)}$ of its posterior distribution, that is, denoting the parameter $\theta^{(m)}$ and its space as $\Theta^{(m)}$,  
\begin{equation}
    \mZ^{(m)}_t = \int_{\Theta^{(m)}}  p(y_{0:t}\mid \theta) \pi_{\theta^{(m)}}(d\theta) = p(y_{0:t} \mid \mM^{(m)}).  
    \label{eq:evidence}
\end{equation}
The model evidence, which was the normalizing constant in Eq.~\eqref{eq:posterior}, can be understood as the density of the observations $y_{0:t}$
given the model $\mM^{(m)}$.
By introducing prior probabilities on the discrete set of model labels 
$\{\mM^{(1)},\ldots, \mM^{(M)}\}$, one can then consider posterior
probabilities of the models given the data, obtained again by Bayes rule as
\begin{align}
\p\left( \mM = \mM^{(m)} \mid y_{0:t} \right) &= \frac{ \p\left( \mM = \mM^{(m)} \right) \mZ_t^{(m)}}{\sum_{m'=1}^M \p\left( \mM = \mM^{(m')}  \right) \mZ_t^{(m')}}.
\label{eq:modelproba}
\end{align}
On top of parameter uncertainty, model uncertainty can be taken into account when performing filtering, prediction or smoothing.
For instance the predictive distribution $\mathcal{P}(dy_{t+k}\mid y_{0:t})$ of
new observations $y_{t+k}$ given $y_{0:t}$ under both model and parameter
uncertainty can be written as
\begin{align}
    &\int_\Y \varphi(y_{t+k}) \; \mathcal{P}(dy_{t+k}\mid y_{0:t}) \nonumber\\
    &=  \sum_{m=1}^M \int_{\Theta^{(m)}} \int_\Y \varphi(y_{t+k}) \;  
    \underbrace{p(dy_{t+k}\mid y_{0:t}, \theta,\mM^{(m)})}_{\text{predictive distribution of y}}
\underbrace{\pi_{\theta^{(m)},t}(d\theta)}_{\text{posterior on parameter}}
   \underbrace{\p\left( \mM = \mM^{(m)} \mid y_{0:t} \right)}_{\text{posterior on model}}  .
    \label{eq:pred:uncertainty}
\end{align}
This type of quantity is also referred to as model averaging \cite{hoeting1999bayesian}.
Sometimes the task consists in selecting one model among the $M$ proposed ones. Then a standard procedure
is to estimate the posterior odds of model $\mM^{(m)}$ versus model $\mM^{(m')}$:
\begin{align}
    \frac{\p\left( \mM = \mM^{(m)} \mid y_{0:t} \right)}{\p\left( \mM = \mM^{(m')} \mid y_{0:t} \right)}
    &= \underbrace{\frac{p\left( y_{0:t}\mid \mM^{(m)} \right)}{p\left( y_{0:t} \mid \mM^{(m')} \right)}}_{\text{Bayes factor}}
    \times 
    \underbrace{\frac{\p\left( \mM = \mM^{(m)} \right)}{\p\left( \mM = \mM^{(m')} \right)} }_{\text{prior odds}}.
    \label{eq:bayesfactor}
\end{align}
Typically the value $1$ is assigned to the prior odds, corresponding to uniform prior probabilities on the model labels, and thus the posterior odds correspond
to the Bayes factor \cite{kass1995bayes}. It is well-known \cite{mackay1992bayesian} that the Bayes factor embodies the principle of Occam's razor:
``simple'' models (i.e. with a low-dimensional parameter space) are favoured over ``complex'' models (i.e. with a high-dimensional parameter space) 
until enough data have accrued to support the additional complexity.

As discussed in \cite{bernardo2009bayesian}, it can be confusing to specify non-zero prior probabilities on the event ``model $\mM^{(m)}$ is the true, data-generating model'', for each $m\in\{1,\ldots,M\}$.
Often, we do not believe that any of the $M$ candidate models is the true data-generating model. In that case, we can bypass the specification of probabilities of the models being true,
and interpret the model evidence in Eq.~\eqref{eq:evidence} as a prior predictive criterion \cite{dawid1984present}, representing how likely the observations are under the model. The logarithm
of the Bayes factor is then the difference between expected utilities associated to each model, under the logarithmic scoring rule, for the task of predicting observations using the prior \cite{kass1995bayes}. Thus,
even in the setting where each candidate model is mis-specified compared to the data-generating process, the use of Bayes factors is still defensible.

To summarise, there are three layers of uncertainty in hidden Markov models: the hidden process, the model parameters and the model itself. Ideally each of
these uncertainties should be taken into account. 

\subsection{Implicit models \label{sec:sub:implicit}}

The integrals presented above are in general impossible to evaluate,
except for linear Gaussian models. In these models the latent Markov chain is an autoregressive process
and the observations are Gaussian measurements of it. Formally
$x_0 \sim \mathcal{N}(\mu_0,\Sigma_0)$, the Gaussian distribution with mean $\mu_0$ and variance $\Sigma_0$, and for each $t\geq 1$,
\begin{align*}
    x_t = A x_{t-1} + v_t\quad \mbox{ and } \quad y_{t-1} &= B x_{t-1} + w_{t-1},
\end{align*}
where $v_t$ and $w_t$ are Gaussian random variables with zero mean and
variances $\Sigma_x$ and $\Sigma_y$ respectively. The parameter is thus $\theta = (\mu_0,\Sigma_0,A,\Sigma_x, B, \Sigma_y)$ and is made
either of real values or of vectors and matrices of compatible dimensions. The
linearity and Gaussianity of the model equations imply that various conditional
distributions of interest, for instance the filtering distribution of $x_t$
given $y_{0:t}$ and $\theta$, are also Gaussian. The Kalman filter
\cite{anderson2012optimal} provides the mean and variance of these Gaussian
distributions for a computational cost of $\mO(t)$.  As a by-product the
likelihood in Eq.~\eqref{eq:likelihood} can be evaluated for any $\theta$ for a
cost in $\mO(t)$, which allows parameter estimation using standard
techniques, sometimes under the name of system parameter identification, see
e.g. \cite{solo1989adaptive}. In a nutshell, linearity and Gaussianity make integration with 
respect to the hidden Markov chain analytically possible and thus 
the integrals of Section~\ref{sec:sub:objects} can be either evaluated or 
at least approximated using standard techniques such as Markov chain Monte Carlo.

As discussed in \cite{papaspiliopoulos2013optimal} there are some
non-Gaussian models for which the filtering
distribution is tractable.  However it has been argued, see e.g. \cite{Breto}, that models
should preferably be proposed based on scientific grounds rather than on
computational ones.  In practice scientists often come up with complex and
generative models, and then use linearization and Gaussian assumptions only to
enjoy the numerical efficiency of Kalman filters and related methods (e.g.
Extended Kalman filter, Ensemble Kalman filter), particularly in
high-dimensional settings \cite{frei2013bridging}. Linearization and
Gaussianity assumptions result in a systematic bias in the subsequent estimation, compared to the
results that would be obtained under the original model.

Let us have a look at a simple generative model that has
been proposed to model the interaction of phytoplankton and zooplankton 
in \cite{jones2010bayesian,murray2013disturbance}. 
The ``PZ'' model is a variation of a Lotka-Volterra model for interactions between prey and predator. 
Phytoplankton are modeled as prey on which zooplankton are grazing. Over successive days indexed by $t$, the model
describes the stochastic growth rate of prey $\alpha_t$,
the population size of prey $p_t$ and the population size of predator $z_t$.
Thus the hidden state $x_t = (\alpha_t, p_t, z_t)$ is three-dimensional.
The stochastic growth rate $\alpha_t$ follows the same distribution every day $t\geq 0$: $\alpha_t \sim \mathcal{N}(\mu_\alpha,\sigma_\alpha^2)$.
The initial distributions for both species are given by
\[\log p_0  \sim \mathcal{N}(\mu_p, \sigma^2_p) \quad \text{and}\quad \log z_0  \sim \mathcal{N}(\mu_z, \sigma^2_z).\]
The transition of $(p_t)$ and $(z_t)$ is jointly described by the differential equation
\begin{align*}
    \frac{dp_t}{dt} &= \alpha p_t - c p_t z_t ,\\
    \frac{dz_t}{dt} &= e c p_t z_t -m_l z_t -m_q z_t^2.
\end{align*}
This is to be interpreted as follows: given a value for $p_{t-1}$ and
$z_{t-1}$, and a draw $\alpha$ of $\alpha_t$, the next states $p_t$ and $z_t$
are obtained as the deterministic solution of the above equation over one time
unit. 
In the equation, $c$ represents the clearance rate of the prey, $e$ is the
growth efficiency of the predator, $m_l$ and $m_q$ are the linear and quadratic
mortality rates of the predator.  
Note that $(p_t,z_t)$ can also be defined at non-integer times, and we
could consider $\alpha_t$ to be a piecewise constant continuous time process
jumping at each integer time according to $\mN(\mu_\alpha, \sigma_\alpha^2)$.
However the observations are gathered in discrete time and thus we find it more
convenient to specify the hidden Markov chain in discrete time
as well. To summarize, given $x_{t-1} = (\alpha_{t-1}, p_{t-1}, z_{t-1})$,
the next state is obtained by drawing $\alpha_t$ from $\mN(\mu_\alpha, \sigma_\alpha^2)$
and $(p_t,z_t)$ is the solution of the ordinary differential
equation above. In practice, this solution can be approximated with arbitrary precision by numerical solvers such as
Runge--Kutta methods.  Note also that the
difference between the PZ model and the classical Lotka-Volterra lies in the
addition of a quadratic mortality term, and in the randomness of the growth
rate $\alpha_t$ sampled at each integer time. 

Given the Markov process $(x_t)_{t\in\N}$, the observations
are noisy measurement of the phytoplankton, $\log y_t \sim \mathcal{N}(\log
p_t, \sigma_y^2)$; the zooplankton are not measured. Indeed it is comparatively
easier to measure the concentration of phytoplankton in a volume of water due
to the fluorescence of the chlorophyll that they contain.  For simplicity we
set $c = 0.25$ and $e = 0.3$, $\mu_p = \mu_z = \log 2$, $\sigma_p = 0.2$,
$\sigma_z = 0.1$, and leave the remaining constants as unknown (or ``free'')
parameters. Thus we introduce  $\theta = (\mu_\alpha, \sigma_\alpha, \sigma_y,
m_l, m_q)$.  Figure~\ref{fig:pz:obs} was obtained by generating a year of data
from the model, with parameters $\mu_\alpha = 0.7, \sigma_\alpha = 0.5,
\sigma_y = 0.2, m_l = 0.1, m_q = 0.1$.  A prior distribution is put on the
model parameters, simply chosen to be a uniform distribution on $[0,1]$ for
each of the five components.  For this application the interest lies both in
the prediction of future states and in parameter inference.

The PZ model is very standard from a scientific point of view as Lotka-Volterra
equations date back to the 1930s (see the historical introduction in
\cite{berryman1992}).  However, from a statistical point of view the model is
not linear nor Gaussian and thus the integrals defined in Section
\ref{sec:sub:objects} are impossible to evaluate exactly.  The model is
generative in the sense that trajectories $x_{0:t}$ of the hidden Markov chain
can be sampled, if not exactly, at least with arbitrary precision using numerical solvers of ordinary differential equations,
for any parameter $\theta$.  Then a series of observations $y_{0:t}$ can be simulated given a
path $x_{0:t}$ and $\theta$.  When the transition distribution $f$ is such that
$x_t$ can be sampled given $x_{t-1}$ and $\theta$ but its transition density
cannot be evaluated, the model is said to be ``implicit'' \cite{Breto}. Here
the ability to evaluate the transition density $f(x_t\mid x_{t-1},\theta)$ for
a given triplet $(x_t,x_{t-1},\theta)$ would in fact depend on the numerical
solver being used, but in general we do not want to assume that we know how to
perform this computation.

Aside from Markov chains defined by differential equations, another scenario where sampling from the transition distribution is easier than
evaluating its probability density function occurs when the transition involves
latent variables. An example is the L\'evy driven stochastic volatility model
described in \cite{bns:real}. Given the previous state $x_{t-1}$ and a
parameter $\theta$, $x_t$ is obtained by sampling an integer-valued random
variable $k$, and then $k$ other random variables $v_{1:k}$ independently from
some distribution $p(dv\mid\theta)$.  The state $x_t$ is obtained as $x_{t} =
\psi(x_{t-1}, k, v_{1:k})$ for some deterministic function $\psi$ that can be
evaluated point-wise.  Thus $x_t$ is straightforward to simulate, but the
evaluation of its transition density involves an integral of $\psi$ over $k$
and $v_{1:k}$, which is not analytically available for general functions $\psi$
and random variables $(k,v_{1:k})$.

\subsection{Online and exact inference \label{sec:sub:blackbox}}

As mentioned earlier, the integrals of Section~\ref{sec:sub:objects} are
impossible to evaluate exactly for general hidden Markov models such as the PZ model 
described in Section~\ref{sec:sub:implicit}. Let us generically denote by $I_t$ one of these integrals,
for instance the one in Eq.~\eqref{eq:filtering:uncertainty}.
Monte Carlo methods have been actively  
developed for hidden Markov models inside and outside the Bayesian framework \cite{CapMouRyd,RobCas},
and yield a random variable $\widehat{I}_t$ estimating $I_t$.
Let us list some desirable features of $\widehat{I}_t$ and of the algorithm producing it, 
in the context of time series.

Numerical methods are said to be ``exact'' if, at any time $t$ they produce consistent estimators 
of the integral $I_t$, where consistency
is with respect to a tuning parameter $N$ of the algorithm producing the estimator. For instance, if $\widehat{I}_t$ converges to $I_t$
when $N$ goes to infinity in the $L_2$ or ``mean square'' sense, then both the bias and the variance of the estimator go to zero, and the algorithm is considered exact.
On the contrary the use of Extended Kalman filters for non-linear non-Gaussian models
results in a systematic bias that cannot be reduced to zero, and this bias is typically hard to quantify. 
More trivially, the estimator that always returns ``one'' has zero variance, a bias that is uniformly bounded over the time index $t$
(if $I_t$ is so itself); but it does not have an algorithmic parameter $N$ allowing a trade-off between computational power 
and precision.

Numerical methods for time series are ``sequential''
if an already obtained estimator can be ``updated'' upon the arrival
of a new observation. For instance if an estimator $\widehat{I}_t$ 
of $I_t$ has been obtained at time $t$, then a sequential method yields $\widehat{I}_{t+1}$ once $y_{t+1}$ is made available,
for a computational cost independent of $t$. The advantage of sequential methods for time series
will be illustrated in Section~\ref{sec:numerics}. A sequential method is said to be ``online'' only 
if its performance does not deteriorate with $t$. Introducing the relative
error $r(\widehat{I}_t)$ of the estimator, typically defined as
\[r(\widehat{I}_t) = \frac{\left(\mathbb{E}\left[(\widehat{I}_t - I_t)^2\right]\right)^{1/2}}{\vert I_t\vert },\]
then the method is online if $r(\widehat{I}_t)$ is uniformly bounded from above over the time index $t$.
This requirement rules out 
the use of standard Monte Carlo algorithms for hidden Markov models. For instance, in the case
of Eq.~\eqref{eq:filtering:uncertainty}, a trivial sequential importance 
sampling estimator would sample $N$ draws from the ``prior'' distribution $\pi_\theta(d\theta)p(dx_{0:t}\mid\theta)$ and use
the conditional density $p(y_{0:t}\mid x_{0:t},\theta)$ as an importance weight. 
The resulting estimator can be updated for a fixed cost per observation and yields a consistent answer when $N$ goes to infinity. 
However its variance would typically grow exponentially with $t$, and thus sequential importance sampling is not online
for this problem.

In the next section we review existing sequential and exact Monte Carlo methods to approximate the
objects described in Section~\ref{sec:sub:objects}, and we discuss whether or
not they satisfy the above online requirement.

\section{Plug and play methods \label{sec:methods}}

Because of the numerous examples of implicit models
as the PZ model of Section~\ref{sec:sub:implicit} we focus on numerical methods that are
compatible with implicit models. These methods, 
called ``plug and play'' (or ``equation free'') in \cite{Breto},
require only the ability to sample
the hidden Markov chain, and to evaluate the measurement density.

\subsection{Approximate Bayesian Computation \label{sec:sub:systematicerror}}

Perhaps the most natural plug and play algorithm for implicit models is the ABC (Approximate Bayesian Computation) method
\cite{toni2009ABCdynamical,beaumont2010approximate,dean2014parameter}. In a nutshell, ABC draws approximately from the posterior distribution of $(\theta,x_{0:t})$ given $y_{0:t}$ using 
the following steps.
\begin{enumerate}
    \item Draw $\theta$ from the prior distribution $\pi_\theta$.
    \item Draw $x_{0:t}$, a realisation of the hidden Markov chain given $\theta$.
    \item Draw $\hat{y}_{0:t}$, a realisation of the observations given $x_{0:t}$ and $\theta$.
    \item If $\mathcal{D}(\hat{y}_{0:t}, y_{0:t})\leq \varepsilon$, retain $(\theta, x_{0:t})$, otherwise go back to step (1).
\end{enumerate}
In this algorithm, $\mathcal{D}$ can be understood loosely as a distance
defined on the observation space $\Y$ \cite{marin2012approximate}.  The value
of $\varepsilon$ has to be chosen by the user; for smaller values, the
synthetic data has to be closer (in the sense of $\mathcal{D}$) to the true
data in order for $(\theta,x_{0:t})$ to be retained. It can be shown that if
$\mathcal{D}$ is a true distance on $\Y$ then when $\varepsilon$ goes to zero,
the procedure samples from the true posterior distribution.  When $\mathcal{D}$
is not a distance (for instance when it is based on summary statitics), or when
$\varepsilon$ is a fixed value, then the samples obtained from ABC are not
distributed according to the posterior distribution, and it is notoriously
difficult to quantify the bias between the obtained approximation and the
target distribution. Thus ABC estimators are typically not exact in the sense of Section~\ref{sec:sub:blackbox}. Note that ABC only requires the ability to sample from $\mu$,
$f$ and $g$.

The next section introduces the particle filter, which is an exact, online and plug and play method
to perform filtering and prediction for a given parameter value. On the other hand, particle filters
require point-wise evaluations of the measurement density $g$, therefore they are less
generally applicable than ABC.

\subsection{Particle filters\label{sec:sub:methods:filtering}}

Particle filters have become the preferred methods to deal with filtering, prediction and smoothing tasks 
\cite{doucet2011tutorial,del2014particle}, since the publication of the seminal
paper \cite{Gordon}. They were initially introduced to evaluate integrals such as Eq.~\eqref{eq:filtering}
for non-linear, non-Gaussian hidden Markov models. The original algorithm, or  
``bootstrap particle filter'', is described in Algorithm~\ref{alg:particlefilter}.

\begin{algorithm}
    \caption{Particle filter\label{alg:particlefilter} for a given parameter $\theta$.}
    \begin{algorithmic}[1]
        \STATE Draw for each $k \in \{1, \ldots, N_x \} \quad x_{0}^{k} \sim \mu(dx_0 \mid \theta)$. \label{alg:line:init}
        \FOR {$t=0$ to $T$}
        \STATE [weighting] Compute for each $k$,  $w_{t}^{k} = g(y_t \mid x_t^k, \theta)$. 
        \STATE [resampling] Sample ancestors $a_t^{1:N_x} \sim r(da^{1:N_x}\mid w_t^{1:N_x})$. \label{alg:line:resampling}
        \STATE [transition] Draw for each $k$, $x_{t+1}^{k} \sim f(dx_{t+1}\mid x_{t}^{a_t^k}, \theta)$.\label{alg:line:transition}
        \ENDFOR
    \end{algorithmic}
\end{algorithm}

The algorithm requires the user to specify a number of ``particles'' $N_x\in\N$. The transition and weighting steps propagate
the samples from one distribution to the next following standard importance sampling \cite{RobCas}.
The resampling step consists in selecting the particles according to their weights, so that the particles
with lowest weights are killed while the particles with highest weights get replicated and propagated. Without the resampling step, 
only one particle would have a significant weight after a few time steps. Various resampling schemes exist
as described in \cite{doucet2011tutorial}.
A standard resampling scheme consists in drawing each ancestor $a^k_t$ 
independently from a categorical distribution with parameters $w_{t}^{1:N_x}$, so that for each $i\in\{1,\ldots,N_x\}$,
$\p(a^k_t = i) = w_{t}^i / \sum_l w_{t}^l$. In Algorithm~\ref{alg:particlefilter}, the resampling distribution is
denoted by $r(da^{1:N_x}\mid w_t^{1:N_x})$ on Line \ref{alg:line:resampling}.

The algorithm yields 
at each step $t$ a weighted sample $(x_t^k, w_t^k)_{k=1}^{N_x}$ approximating the filtering
distribution $p(dx_t\mid y_{0:t}, \theta)$, in the sense that the integral in Eq.~\eqref{eq:filtering}
can be approximated by the weighted average
\begin{equation}
    \frac{1}{\sum_{k=1}^{N_x} w_t^k} \sum_{k=1}^{N_x} w_t^k \varphi(x_t^k)
    \label{eq:filtering:estimator}
\end{equation}
in a consistent manner as $N_x$ goes to infinity. A rich theoretical literature
supports particle filters. Some of the important results state that the
estimator in Eq.~\eqref{eq:filtering:estimator} satisfies a central limit
theorem (CLT), that both the bias and the variance are of order $1/N_x$, and that for a
finite number $N_x$, the variance of the estimator can be uniformly bounded over
the time steps $t$
\cite{Chopin:CLT,delMoral:book,del2014particle,whiteley2013stability}. The CLT
and the time uniform results are remarkable since they make the particle filter
``online'' and ``exact'' in the sense of Section~\ref{sec:sub:blackbox}.  Particle filters
have therefore become standard tools for filtering in hidden Markov models. Kalman filter
techniques such as Ensemble Kalman Filters are still used when the dimension $d_x$ of the state space $\X$ is
large \cite{frei2013bridging}, because the variance of the particle filter estimators typically grows
exponentially with $d_x$. Algorithmic improvements and theoretical studies of the
impact of the dimension on particle filters have been recently proposed in
\cite{rebeschini2013can,bertoli2014adaptively,beskos2014stable}. Exact and online methods for large dimensional filtering
problems constitute an active area of research.

An important by-product of this algorithm is an estimator $\widehat{Z}_t(\theta)$ of
the marginal likelihood of $\theta$ at time $t$, as defined in Eq.~\eqref{eq:likelihood}. The estimator takes the simple form
\begin{equation}
    \widehat{Z}_t(\theta) = \prod_{s=0}^t \left(\frac{1}{N_x} \sum_{k=1}^{N_x} w_s^k\right),
    \label{eq:likelihood:estimator}
\end{equation}
and can thus be updated sequentially at each step of Algorithm
\ref{alg:particlefilter}.  This estimator has also been extensively studied in
the literature. It happens to be unbiased, and by scaling the number of particles
$N_x$ linearly with the number of observations $t$, its relative variance is bounded by a constant,
as proved in \cite{CerDelGuy2011nonasymptotic}. Hence the estimation of the likelihood
using particle filters is not ``online'': for a fixed cost per observation, the relative
error goes to infinity. The cost of estimating the likelihood $p(y_{0:t}\mid \theta)$ can be said to be quadratic 
in the sense that one needs to choose $N_x = t$ to guarantee a fixed relative error. Then the 
cost of running a particle filter with $N_x$ particles for $t$ steps is $t^2$, in the number of evaluations of $g$ and
draws from $f$.

To perform filtering on the path space or smoothing, one can simply modify
Algorithm~\ref{alg:particlefilter} to keep track of the generated paths
$x_{0:t}^k$ instead of the most recent states $x_t^k$, as in Section 3.5 of
\cite{doucet2011tutorial}. Thus one would define $\bar{x}_0^k = x_0^k$ on Line
\ref{alg:line:init} of Algorithm~\ref{alg:particlefilter}, and then on Line
\ref{alg:line:transition} one would define $\bar{x}^k_{0:t+1} =
(\bar{x}_{0:t}^{a_t^k},x^k_{t+1})$.  The resulting ``path particles''
$(\bar{x}_{0:t}^k, w_t^k)_{k=1}^{N_x}$ approximate the filtering distribution
on the full path $p(dx_{0:t}\mid y_{0:t}, \theta)$, mentioned in Section
\ref{sec:sub:objects}. Thus, the estimator 
\begin{equation*}
    \frac{1}{\sum_{k=1}^{N_x} w_t^k} \sum_{k=1}^{N_x} w_t^k \varphi(\bar{x}_{0:t}^k)
\end{equation*}
converges to the integral of $\varphi$ with respect to
$p(dx_{0:t}\mid y_{0:t}, \theta)$ when ${N_x}$ goes to infinity and satisfies a
CLT (Chapter 9 of \cite{delMoral:book}).  However when $t$
increases the variance of this estimator quickly deteriorates due to the
well-known path degeneracy phenomenon. The variance has been shown to increase
at least quadratically, and in general exponentially, as a function of $t$ in
\cite{del2003class,DoucPoyiaSin}.  Indeed the resampling steps prune the
population of distinct path particles at each time step. Let us denote by $\bar{x}_{0:t}^k(s)$,
for $0\leq s\leq t$, the $s$-th component of a path $\bar{x}_{0:t}^k$. Then at a given time
$t$, the latest components $\bar{x}_{0:t}^{1:N_x}(t)$ of the path particles
$\bar{x}_{0:t}^{1:N_x}$ are all distinct, but the first components $\bar{x}_{0:t}^{1:N_x}(0)$
contain many replicate values. In fact the number of unique values in
$\bar{x}_{0:t}^{1:N_x}(0)$ quickly decreases to only one as $t$ increases.  More precisely the
number of unique elements among the $N_x\times (t+1)$ components that compose the
$N_x$ path particles $\bar{x}_{0:t}^{1:N_x}$ has been upper bounded by $(t+1) + C N_x
\log N_x$ in expectation in \cite{jacob2013path}, where $C$ is independent of $t$ and $N_x$. To resolve the path degeneracy
issue for the problem of smoothing given a parameter value $\theta$, many
particle algorithms have been proposed, such as fixed-lag approximations, forward filtering backward
smoothing or two filter formula, as described in
\cite{douc2011,doucet2011tutorial}. However the path degeneracy phenomenon also has
consequences on parameter estimation, as described in the next section.

\subsection{Particle-based approaches to parameter estimation \label{sec:sub:methods:parameter}}

The early attempts to estimate the parameters using particle methods involve a
reparametrization where the parameters $\theta$ are treated as an extra
component of the hidden states.  Thus a new hidden Markov model is introduced,
where the hidden state is $\tilde{x}_t = (x_t, \theta_t)$ for all times $t$,
$x_t$ being the hidden state of the original model.  The new initial
distribution is then $\pi_\theta(d\theta_0) \mu(dx_0\mid \theta_0)$, where
$\pi_\theta$ is the prior on the parameters of the original model, and $\mu$
the original initial distribution.  The new transition is
$\delta_{\theta_{t-1}}(d\theta_t) f(dx_t\mid x_{t-1},\theta_{t})$, where 
$\delta_x$ represents the Dirac measure centered at the point $x$ and 
$f$ is the transition of original model. Finally the new measurement distribution
is defined as $g(dy_t \mid x_t, \theta_t)$.  Then by performing filtering on
the modified model, one obtains a particle approximation of the distribution of
$\tilde{x}_t$ given $y_{0:t}$, and the distribution $\pi_{\theta,t}(d\theta)$
of $\theta$ given $y_{0:t}$ is obtained as a marginal distribution thereof.

The idea traces back to \cite{Kitagawa:Self}, who already recognized the
occurrence of path degeneracy. Indeed, the parameter values $\theta_t^{1:N_x}$ are resampled as
part of the states, but contrary to the states $x_t^{1:N_x}$ they are never diversified for the transition of the
parameters is a delta function. Hence there are fewer and fewer unique values in
the particle approximation of the posterior distribution
$\pi_{\theta,t}(d\theta)$. We recognize the similarity with particle filtering on the path space, as described in the previous section.
Early attempts such as \cite{Kitagawa:Self,LiuWest} 
proposed to replace the delta function by a Gaussian random walk in order to introduce diversity among the parameter samples.
Alternatively, it has been proposed to introduce Markov chain Monte Carlo (MCMC) moves to diversify the parameter values \cite{GilksBerzu,fearnhead2002markov}.
Those moves have the benefit of leaving the correct posterior distribution invariant. 
However the high dimensionality of $(\theta,x_{0:t})$ makes the design of efficient Markov chain Monte Carlo moves challenging in the setting of hidden Markov models,
as well as the high correlation between the parameters and the states and between consecutive states \cite{pap:rob:sk}.
Finally \cite{fearnhead2002markov,Storvik} proposed specific moves in models
such that the distribution of $\theta$ given $x_{0:t}$ and $y_{0:t}$ only
depends on $x_{0:t}$ through a low dimensional sufficient statistic; see the
early criticism in \cite{Andrieu05onlineparameter}. Reviews of various
parameter estimation methods are proposed in \cite{kantas2009overview,kantas2014overview}.

The inefficiency of standard MCMC moves and the path degeneracy phenomenon have
made the various attempts at estimating the parameters as part of the hidden
states generally unsuccessful. In the recent years two major advances have been
proposed based on particle filters. The first one is Iterated filtering
\cite{ionides2006inference,ionides2011iterated}, which is an optimization
method relying on particle filters and an original representation of the score
to find the maximum likelihood estimator in implicit models. The second one is
particle Markov chain Monte Carlo \cite{PMCMC}, a class of MCMC algorithms
using particle filters to design efficient proposal distributions on the space
of $(\theta,x_{0:t})$. Here we recall a particle MCMC method
called particle marginal Metropolis-Hastings (PMMH). 

The pseudo-code is given in Algorithm~\ref{alg:pmmh}.  As described in Section
\ref{sec:sub:methods:filtering}, particle filters can be defined on the path
space and thus yield a sample $(\bar{x}_{0:t}^k, w_t^k)_{k=1}^{N_x}$ of
trajectories approximating the distribution of the paths given the observations
and the parameter. In the pseudo-code, ``drawing a path'' means randomly selecting one
of these paths with probabilities proportional to their weights $w_t^{1:N_x}$.
Extracting $\widehat{Z}_t$ means computing the estimator of Eq.~\eqref{eq:likelihood:estimator}. The proposal distribution $q(d\theta^\star
\mid \theta)$ can be a Gaussian distribution centered on
$\theta$.  Intuitively, if the number $N_x$ of particles was infinite, then
drawing a path among the path particles would be equivalent to perfect sampling 
from the filtering distribution $p(dx_{0:t}\mid y_{0:t},\theta^\star)$, and
the likelihood estimator $\widehat{Z}_t(\theta^\star)$ would yield a perfect
evaluation of the likelihood $p(y_{0:t}\mid\theta^\star)$. Thus the algorithm
would be a standard Metropolis-Hastings with proposal distribution
$q(d\theta^\star\mid \theta)p(dx_{0:t}^\star \mid y_{0:t}, \theta^\star)$ and
target distribution $\pi_{\theta,t}(d\theta)p(dx_{0:t}\mid y_{0:t}, \theta)$.
The remarkable result of \cite{PMCMC} is that for any finite $N_x$, the PMMH
algorithm also generates a Markov chain with invariant distribution
$\pi_{\theta,t}(d\theta)p(dx_{0:t}\mid y_{0:t}, \theta)$. Unsurprisingly
larger values of $N_x$ yield better performance of the algorithm and convergence
properties of particle MCMC methods have been studied in
\cite{andrieu2012convergence,andrieu2014establishing,andrieu2013uniform,chopin2013particle,lindsten2014uniform}.
In terms of computational cost, the number of particles $N_x$ has to be chosen
proportional to $t$ in order to control the variance of the likelihood
estimator. Thus each step of PMMH costs $t^2$, as was conjectured in \cite{PMCMC} and more formally
studied in \cite{pittdoucet2015,sherlock2015}. It is less clear how the number
of iterations $N_\theta$ must be chosen as a function of the number of
observations $t$, although some results obtained for standard MCMC 
could be informative \cite{belloni2009computational}. Under the rather
optimistic assumption that $N_\theta$ can be chosen independently of $t$, then
the algorithm would overall be of quadratic cost with respect to the number of
observations.

\begin{algorithm}
    \caption{Particle marginal Metropolis-Hastings.\label{alg:pmmh}}
  \begin{algorithmic}[1]
      \STATE Set some $\theta^{(1)}$.
      \STATE Run a particle filter with $N_x$ particles given $\theta^{(1)}$.
      \STATE Extract $\widehat{Z}_t(\theta^{(1)})$ and draw one path $x_{0:t}^{(1)}$.
      \FOR {$i=2$ to $N_\theta$}
      \STATE Propose $\theta^\star \sim q(d\theta^\star | \theta^{(i-1)})$.
      \STATE Run a particle filter with $N_x$ particles given $\theta^{\star}$.
      \STATE Extract $\widehat{Z}_t(\theta^{\star})$ and draw one path $x_{0:t}^{\star}$.
      \STATE Compute:
      \[\alpha = \text{min}\left(
          1, \frac{\widehat{Z}_t(\theta^\star)p(\theta^{\star})}{\widehat{Z}_t(\theta^{(i-1)})p(\theta^{(i-1)})}\frac{q(\theta^{(i-1)}|\theta^{\star})}{q(\theta^{\star}|\theta^{(i-1)})}
      \right).
      \] 
      \STATE Set $(\theta^{(i)} , x_{0:t}^{(i)}) = \begin{cases}
                   (\theta^{\star}, x_{0:t}^{\star}) \quad \text{with probability }\alpha,\\
                   (\theta^{(i-1)}, x_{0:t}^{(i-1)}) \quad \text{with probability } 1 - \alpha .
                  \end{cases}$
%       \STATE Set $\theta^{(i)} = \theta^{\star}$, $x_{0:t}^{(i)} = x_{0:t}^{\star}$ with probability $\alpha$.
%       \STATE Otherwise set $\theta^{(i)} = \theta^{(i-1)}$, $x_{0:t}^{(i)} = x_{0:t}^{(i-1)}$.
      \ENDFOR
  \end{algorithmic}
\end{algorithm}

Various techniques can be used to process the output of MCMC methods to compute
estimators of the model evidence as in Eq.~\eqref{eq:evidence}
\cite{carlin1995bayesian,gelman1998simulating}. Evidence estimators based on particle MCMC outputs have been proposed in
\cite{peters2010ecological}. Thus particle MCMC methods constitute the first
class of methods providing practical and consistent approximations of the
objects of interest mentioned in Section~\ref{sec:sub:objects} in the context
of general implicit models.  By design they are iterating over the full dataset
$y_{0:t}$, and thus constitute ``offline'' or ``batch'' methods, as opposed to the sequential and online features 
described in Section~\ref{sec:sub:blackbox}. 
Upon the arrival of a new observation $y_{t+1}$, the algorithm has to be run again from the beginning. 
The result of a previous run given $y_{0:t}$ might only be used to design the proposal
distribution $q(d\theta^\star\mid \theta)$ and to choose the number of particles $N_x$.

The SMC$^2$ algorithm has been introduced to address this issue
\cite{chopin2013smc2,fulop2013efficient}, and to take one step towards exact, online plug and play methods for Bayesian inference
in implicit models.
The method processes the observations one after the other, and provides at each
step an updated estimator of the various quantities of interest such as the
ones described in Section~\ref{sec:sub:objects}.

\section{A sequential plug and play algorithm \label{sec:smcsamplers}}

In the light of particle MCMC methods, which mimick the behaviour of ideal MCMC methods when $N_x$ goes to infinity, 
the idea of SMC$^2$ is to mimick an ideal sequential Monte Carlo (SMC) sampler \cite{Chopin:IBIS,DelDouJas:SMC} in the setting of hidden Markov models.

\subsection{SMC samplers \label{sec:sub:smcsamplers}}

We first describe the SMC sampler algorithm that we would like to imitate in the hidden Markov model setting. It has been
originally proposed for simpler models where it is possible to evaluate
the incremental likelihood functions $p(y_t \mid y_{0:t-1}, \theta)$,
for all $\theta$ and all $t$. This is typically the case in parametric models for independent
observations, where $p(y_t \mid y_{0:t-1}, \theta) = p(y_t \mid \theta)$.
Since the full likelihood can be expressed as a product of
those incremental likelihoods, it can be evaluated point-wise,
and thus the standard Metropolis-Hastings (MH) algorithm is applicable in this context to sample
from the posterior distribution $\pi_{\theta,t}(d\theta)$ for a fixed dataset $y_{0:t}$. 
SMC samplers approximate each posterior distribution $\pi_{\theta,t}(d\theta)$
sequentially over the time $t$, that is, upon the arrival of new pieces
of information. The notion of time can be purely artificial here, e.g. when the data correspond
to measurements of different individuals. An adaptive SMC sampler
is described in Algorithm~\ref{alg:Adaptive-SMC-sampler}.
It produces a weighted sample $(\theta_t^k,\omega_t^k)_{k=1}^{N_\theta}$ that approximates
the posterior distribution $\pi_{\theta,t}(d\theta)$ at each time $t$.

\begin{algorithm}
    \caption{Adaptive SMC sampler. \label{alg:Adaptive-SMC-sampler}}
    \begin{algorithmic}[1]
        \STATE Draw for each $k \in \{1, \ldots, N_\theta\}\quad\theta_{0}^{k} \sim \pi_\theta(d\theta)$.
        \STATE Set for each $k$, $\omega_{-1}^{k}=N_\theta^{-1}$.
        \FOR {$t=0$ to $T$}
        \IF {ESS($\omega_{t-1}^{1:N_\theta})<c$}
        \STATE Construct a proposal distribution $q_{t-1}$ based on the particles $(\theta^k_{t-1},\omega_{t-1}^k)_{k=1}^{N_\theta}$.
        \STATE [resampling] Sample ancestors $a_t^{1:N_\theta}\sim r(da^{1:N_\theta}\mid w_{t-1}^{1:N_\theta})$.
        \STATE For each $k$, replace $\theta_{t-1}^k$ by $\theta_{t-1}^{a_t^k}$.
        \STATE Set for each $k$, $\omega_{t-1}^{k}=N_\theta^{-1}$.
        \STATE [move] Perform an MCMC move on each particle $\theta_{t-1}^k$ using $q_{t-1}$, leaving $\pi_{\theta,t-1}(d\theta)$ invariant.
        \ENDIF
        \STATE Set for each $k$, $\theta_{t}^{k}=\theta_{t-1}^{k}$.
        \STATE [weighting] Update for each $k$, $W_{t}^{k}=\omega_{t-1}^{k}\; p(y_{t}\mid y_{0:t-1}\theta_{t}^{k})$.
        \STATE Normalize for each $k$, $\omega_{t}^{k}=W_{t}^{i}/\sum_{k=1}^{N_\theta}W_{t}^{k}$.
        \ENDFOR
    \end{algorithmic}
\end{algorithm}

In the algorithm, the resampling step is similar to the one of Algorithm~\ref{alg:particlefilter},
but it is triggered only when the effective
sample size (ESS) falls below a threshold $c$. The ESS is an assessment of the degeneracy of the
weights and takes values between $0$ and $1$. It is defined as the following function of the normalized
weights:
\[
    \text{ESS}\left(\omega^{1:N_\theta}\right)=\frac{1}{N_\theta\sum_{k=1}^{N_\theta}\left(\omega^{k}\right)^{2}}.
\]
This adaptive resampling scheme could be applied to the particle filter of Section~\ref{sec:sub:methods:filtering} as well,
but it proves crucial for SMC samplers, for complexity reasons that will become clear in Section~\ref{sec:sub:smc2complexity}.
The combination of the resampling and the move steps is called the rejuvenation step.

A simple choice of move step leaving $\pi_{\theta,t}(d\theta)$ invariant is to apply a MH kernel with independent proposals from $q_t(d\theta)$.
The proposal distribution can be a Gaussian distribution with
mean and variance taken as the empirical mean and variance of the particles $(\theta_t^k,\omega_t^k)_{k=1}^{N_\theta}$ \cite{Chopin:IBIS}.
Note that under Bernstein-Von Mises
conditions, the posterior distribution converges itself to a Gaussian distribution, and thus using an adaptive Gaussian proposal
distribution in the rejuvenation step is an asymptotically optimal choice.
The move step then consists in applying one step of MH to each of the $N_\theta$ particles.

The algorithmic parameters left to choose are the number of particles $N_\theta$ and the ESS
threshold $c$, which can be set to $50\%$ by default. Higher values mean more rejuvenation steps,
which constitute the bulk of the computational cost of the algorithm. 
The algorithm has been extended to a more general form in \cite{DelDouJas:SMC}, which allows various algorithmic improvements
as well as a unified theoretical study under the Feynman-Kac framework. Various articles 
study its theoretical properties \cite{jasra2008stability,whiteley2012sequential,schweizer2012non}. In particular \cite{schweizer2012multimodal}
demonstrates its theoretical advantage over MCMC when the posterior distribution is multimodal. The effect of triggering resampling steps
based on an ESS criterion has been studied in \cite{delmoral2012}. The behaviour of the algorithm with respect to the dimension $d_\theta$ of the parameter space
has been studied in \cite{beskos2014}.

Similarly to the likelihood estimator given by particle filters in Eq.~\eqref{eq:likelihood:estimator}, SMC samplers yield an estimator
$\widehat{\mZ}_t$
of the evidence $\mZ_t$ defined in Eq.~\eqref{eq:evidence}, that can be computed as
\begin{equation}
    \widehat{\mZ}_t = \prod_{s=0}^t \left(\sum_{k=1}^{N_\theta} \omega_{s-1}^k p(y_s\mid y_{0:s-1},\theta^k_{s})\right).
    \label{eq:evidence:estimator}
\end{equation}
The form of the estimator is slightly different from the likelihood estimator
in Eq.~\eqref{eq:likelihood:estimator} because the resampling steps are not
applied at every step.   One way to justify it is to consider that if
$(\theta_t^k,\omega_{t-1}^k)_{k=1}^{N_\theta}$ is a consistent particle
approximation of $\pi_{\theta,t-1}(d\theta)$, e.g. in probability, and remembering that the weights
$\omega_{t-1}^{1:N_\theta}$ are normalized in the algorithm, then 
\[ \sum_{k=1}^{N_\theta} \omega_{t-1}^k p(y_t \mid y_{0:t-1},\theta^k_{t}) \xrightarrow[N_\theta\to\infty]{\mathbb{P}} \int_\Theta p(y_t\mid y_{0:t-1}, \theta) \pi_{\theta,t-1}(d\theta) = p(y_t \mid y_{0:t-1}).\]
Taking the product over time steps yields an estimator of the full evidence
$p(y_{0:t})$, given the model. The inclusion of SMC samplers into the
Feynmac-Kac framework of general particle methods allows the study of the
properties of this estimator, in particular \cite{DelDouJas:SMC} obtain a
central limit theorem. Empirically
\cite{zhou2013towards} demonstrate the advantage of SMC samplers over MCMC methods to estimate the model evidence.

\subsection{An approximate SMC sampler for hidden Markov models \label{sec:sub:smc2}}

The original SMC sampler as in Algorithm~\ref{alg:Adaptive-SMC-sampler} cannot
be directly applied to the hidden Markov model scenario, in the same way that standard MH could
not be applied: the likelihood function as defined in Eq.~\eqref{eq:likelihood}
cannot be evaluated point-wise, and the incremental likelihoods $p(y_t\mid
y_{0:t-1},\theta)$ cannot either. Mimicking the reasoning behind particle MCMC
methods, particle filters can be used to obtain estimators of those likelihood
terms. For simplicity, we present an SMC sampler algorithm to sample from
$\pi_{\theta,t}(d\theta)$ only, but the same algorithm can be used to sample
from the joint distribution $\pi_{\theta,t}(d\theta)p(dx_{0:t}\mid y_{0:t},
\theta)$, as shown in \cite{chopin2013smc2}.  The article
\cite{fulop2013efficient} essentially proposed the same algorithm
independently, with challenging applications in econometrics.

To each of the $N_\theta$ parameter values produced by the SMC sampler as in
Algorithm~\ref{alg:Adaptive-SMC-sampler}, we thus attach a particle filter with
$N_x$ particles as in Algorithm~\ref{alg:particlefilter}, hence the name
SMC$^2$ evoking those two layers of particle approximations. To avoid confusion
we will talk about $\theta$-particles and $x$-particles, and denote
respectively by $N_\theta$ and $N_x$ their numbers. At any time $t$, each of
the $N_\theta$ $\theta$-particles is indexed by $k$ as in $\theta_t^k$, while
each of the associated $x$-particles is indexed by $n,k$ as in $x_t^{n,k}$. The
method is described in Algorithm~\ref{alg:SMC2}.

\begin{algorithm}
    \caption{SMC$^2$ sampler. \label{alg:SMC2}}
    \begin{algorithmic}[1]
        \STATE [$\theta$-initialization] Draw for each $k \in \{1, \ldots, N_\theta\}\quad\theta_{0}^{k} \sim \pi_\theta(d\theta)$.
        \STATE [$x$-initialization] For each $k$, draw for each $n \in \{1, \ldots, N_x\} \quad x_0^{n,k}\sim \mu(dx_0\mid \theta_0^k)$.
        \STATE Set for each $k$, $\omega_{-1}^{k}=N_\theta^{-1}$.
        \FOR {$t=0$ to $T$}
        \IF {ESS($\omega_{t-1}^{1:N_\theta})<c$}
        \STATE Construct a proposal distribution $q_{t-1}$ based on the $\theta$-particles $(\theta^k_{t-1},\omega_{t-1}^k)_{k=1}^{N_\theta}$.
        \STATE [$\theta$-resampling] Sample ancestors $a_t^{1:N_\theta}\sim r(da^{1:N_\theta}\mid w_{t-1}^{1:N_\theta})$.
        \STATE For each $k$, replace $\theta_{t-1}^k$ by $\theta_{t-1}^{a_t^k}$.
        \STATE Set for each $k$, $\omega_{t-1}^{k}=N_\theta^{-1}$.
        \STATE [$\theta$-move] Perform a PMMH move on each $\theta$-particle $\theta_{t-1}^k$ using $q_{t-1}$, leaving $\pi_{\theta,t-1}(d\theta)$ invariant.
        \ENDIF
        \STATE Set for each $k$, $\theta_{t}^{k}=\theta_{t-1}^{k}$ .
        \STATE [$x$-weighting] Compute for each $k$ and $n$,  $w_{t}^{n,k} = g(y_t \mid x_t^{n,k}, \theta_t^k)$. 
        \STATE [$x$-resampling] Sample for each $k$, $a_t^{1:N_x,k} \sim r(da^{1:N_x}\mid w_t^{1:N_x,k})$.
        \STATE [$x$-transition] Draw for each $k$ and $n$, $x_{t+1}^{n,k} \sim f(dx_{t+1}\mid x_{t}^{a_t^{n,k},k}, \theta_t^k)$.
        \STATE Compute for each $k$, $\hat{p}(y_t\mid y_{0:t-1},\theta_t^k) = N_x^{-1}\sum_{n=1}^{N_x} w_t^{n,k}$.
        \STATE [$\theta$-weighting] Update for each $k$, $W_{t}^{k}=\omega_{t-1}^{k} \; \hat{p}(y_t\mid y_{0:t-1},\theta_t^k)$.
        \STATE Normalize for each $k$, $\omega_{t}^{k}=W_{t}^{i}/\sum_{k=1}^{N_\theta}W_{t}^{k}$.
        \ENDFOR
    \end{algorithmic}
\end{algorithm}

The algorithm follows the structure of Algorithm~\ref{alg:Adaptive-SMC-sampler}, except that each $\theta$-particle is equipped with
a particle filter that is updated at each step. The differences are summarised in the following two points.
\begin{itemize}
    \item At time $t$, the weight of the $\theta$-particle $\theta_t^k$ is
        updated using an estimator $\hat{p}(y_t\mid y_{0:t-1},\theta_t^k)$
        obtained from the associated particle filter, instead of the true
        incremental likelihood $p(y_t\mid y_{0:t-1},\theta_t^k)$.
    \item The move step to rejuvenate the $\theta$-particles relies on particle MCMC instead of MCMC.
\end{itemize}

A more complete description of the algorithm is
given in \cite{chopin2013smc2}. Let us simply mention that SMC$^2$
is a standard SMC sampler targeting an extended distribution which admits
$\pi_{\theta,t}(d\theta)p(dx_{0:t}\mid y_{0:t}, \theta)$ as one of its marginal
distributions, for any number $N_x$.  Hence, the algorithm falls into the class
of exact approximations, similarly to particle MCMC methods. Thus filtering
under parameter uncertainty as defined in Eq.~\eqref{eq:filtering:uncertainty}
can be addressed consistently, for a finite $N_x$ and $N_\theta$ going to infinity, and furthermore the algorithm is sequential by
design. Before turning to its computational complexity, let us mention that the
model evidence can be retrieved with the
following estimator
\begin{equation}
    \widehat{\mZ}_t = \prod_{s=0}^t \left(\sum_{k=1}^{N_\theta} \omega_{s-1}^k \hat{p}(y_s\mid y_{0:s-1},\theta^k_{s})\right).
    \label{eq:evidence:smc2estimator}
\end{equation}
Thus the algorithm can consistently compute integrals such as Eq.~\eqref{eq:pred:uncertainty}.

\subsection{Complexity of SMC samplers\label{sec:sub:smc2complexity}}

Going back to the ideal SMC sampler in Algorithm
\ref{alg:Adaptive-SMC-sampler}, a MH move for each particle $\theta_t^k$ at
time $t$ requires an evaluation of the likelihood $p(y_{0:t}\mid\theta_t^k)$,
which costs $\mO(t)$.  If a rejuvenation step was performed at every step from time $0$ to $t$, the
algorithm would then cost $\mathcal{O}(N_\theta t^{2})$. Fortunately, the ESS
decreases slower and slower, and thus the rejuvenation step occurs less and
less often, hence the cost typically reduces to $\mathcal{O}(N_\theta t)$,
as shown in Theorem 1 of \cite{Chopin:IBIS}. In other words, at each step $t$,
either the assimilation of $y_{t}$ costs $\mathcal{O}(t)$ or $\mathcal{O}(1)$,
whether or not a rejuvenation step is performed, which happens with a probability
decreasing with $t$. Let us denote by $p_{t}$ the probability of a rejuvenation
step occurring at time $t$. If the other operations are of cost $1$ at each step,
then the overall cost $C_{t}$ for the algorithm to reach step $t$ satisfies
\[
   \mathbb{E}\left[C_{t}\right]= \sum_{s=0}^{t}\left(p_{s} \times s + (1 - p_s)\times 1\right) = (t+1) + \sum_{s=0}^t p_s \times (s-1) 
\]
which indeed is linear in $t$ if $p_{t}=\mathcal{O}(1/t)$. This is another
formulation of the result in \cite{Chopin:IBIS}.  Thus the algorithm is online
in the sense described in Section~\ref{sec:sub:blackbox}. Note that the
algorithm requires more and more memory, as a rejuvenation step at time $t$
involves browsing over the past dataset $y_{0:t}$, which thus has to be kept
available. Thus the algorithm is not ``online'' memory-wise but only in terms
of computational cost. 
One can hope that the errors are uniformly bounded over
time for a fixed $N_\theta$ if the rejuvenation steps are performing equally
well across all time steps. This motivates the adaptation of the proposal
distribution $q_t$ in Algorithm~\ref{alg:Adaptive-SMC-sampler}. 

For the SMC$^2$ algorithm of Algorithm~\ref{alg:SMC2}, the same reasoning
applies, motivated by empirical results such as Figure~\ref{fig:pz:cost} to be
described in the next section.  The difference is that, in order to bound the
errors uniformly over time, and for the particle MCMC steps to perform equally
well across all time steps, one needs to increase the number $N_x$ of
$x$-particles with $t$. Scaling $N_x$ linearly with $t$, the cost of running
$N_\theta$ particle filters with $N_x$ $x$-particles for $t$ steps is
$\mathcal{O}(N_\theta t^2)$. Under the same occurrence pattern of rejuvenation
steps, \cite{chopin2013smc2} obtain an overall computational cost in
$\mathcal{O}(N_\theta t^2)$.  The SMC$^2$ algorithm is thus sequential but not
online. Upon the arrival of a new piece of observation, the estimator can be
updated, but one has to increase the computational effort linearly with the
number of observations in order to obtain time uniform guarantees.  A
modification of SMC$^2$ is proposed in  \cite{chopin2013smc2} so that $N_x$ can
be automatically increased along the observations when required. The acceptance
rate of the rejuvenation steps are monitored to assess whether $N_x$ is large
enough at any time $t$. This modification does not make the algorithm online,
but allows the automatic adjustment of the computational cost to guarantee a stable performance
over time.

There currently exists no method to perform
online and exact Bayesian inference for general hidden Markov models
\cite{crisan2013nested,zhou2015biased}, which poses a serious challenge in the presence of
very long time series.  In terms of scaling with the number of particles, SMC
algorithms are very amenable to modern parallel architectures.  The algorithm
is typically of linear complexity in $N_x$ and in $N_\theta$, and most of the computation can
be done in parallel, except for the resampling steps. This has motivated a
series of articles in the recent years, both in the computational literature
\cite{hong2006high,bolic2005resampling,murray2014parallel} and in the
methodological literature
\cite{whiteley2013role,verge2013parallel,chan2014theory}.

\section{Numerical illustration\label{sec:numerics}}

\begin{figure}
    \centering
    \includegraphics[width=0.9\textwidth]{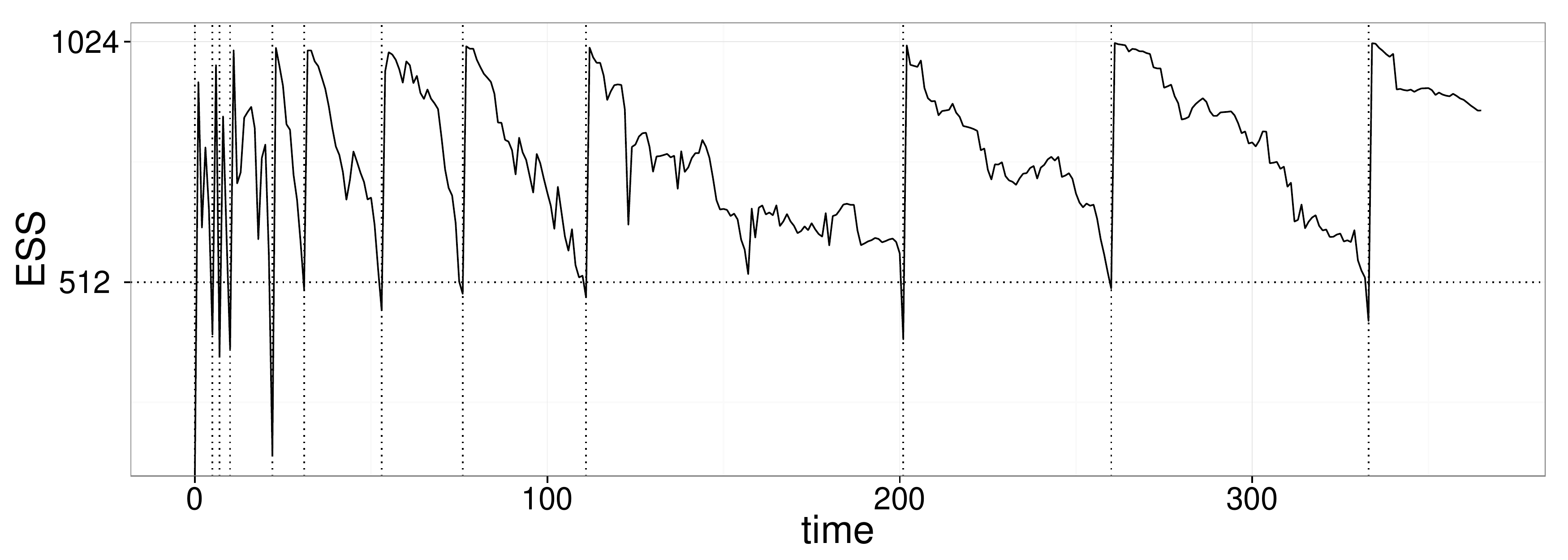}
    \caption{Effective Sample Size against time, over one run of SMC$^2$ on the PZ model. The vertical dashed lines represent the resampling
times and the horizontal dashed line represents the ESS threshold, set to $50\%$ of $N_\theta$.}
    \label{fig:pz:ess}
\end{figure}

The PZ model of Section~\ref{sec:sub:implicit} is used to illustrate the
various outputs of SMC$^2$.  Given the parameters set in
Section~\ref{sec:sub:implicit}, $T = 365$ observations are generated as shown
on Figure~\ref{fig:pz:obs}.  The algorithm is run with $N_\theta = 1024$, $N_x
= 1024$ and an ESS threshold $c$ of $50\%$. The proposal distribution $q_t$ of
the move steps is a Gaussian distribution using the empirical mean and covariance
of the weighted particles $(\theta_t^{k},\omega_t^k)_{k=1}^{N_\theta}$.
Each rejuvenation step performs five successive PMMH moves on each particle
$\theta_t^k$.  The resampling distribution for both the $\theta$-particles and
the $x$-particles is chosen to be the systematic resampling scheme
\cite{CarClifFearn}.  To diagnose the behaviour of an SMC$^2$ run, the
ESS of the $\theta$-particles is plotted against time on Figure~\ref{fig:pz:ess}.  As
expected the ESS decreases slower and slower with the time steps, resulting in
only three rejuvenation steps in the second half of the dataset, whereas ten
rejuvenation steps occurred in the first half. The acceptance rate of the move steps
is found to be above $40\%$ at the end of the run.

\begin{figure}
    \centering
    \includegraphics[width=0.9\textwidth]{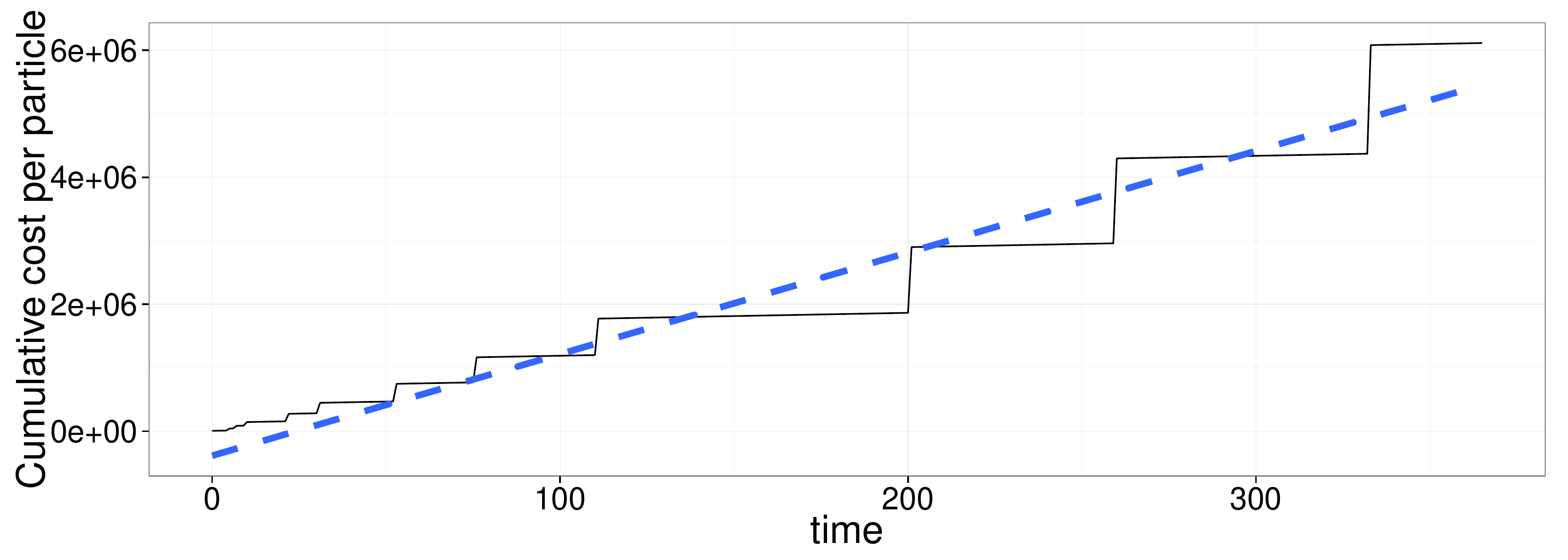}
    \caption{Cumulative cost per $\theta$-particle during one run of SMC$^2$ on the PZ model, with $N_x = 1024$ and five PMMH moves per rejuvenation step.
    The cost is measured in the number of calls to the transition sampling function and the measurement density function. The dashed line represents a linear regression of the cost over
the time index.}
    \label{fig:pz:cost}
\end{figure}

The computational cost of the algorithm is represented on Figure~\ref{fig:pz:cost}. More precisely what is plotted is the number of times that
the transition $f$ and the measurement $g$ are called, for each of the $1024$
$\theta$-particles. Since there are $365$ time steps and $1024$ $x$-particles
per $\theta$-particle, if no PMMH move was performed there would be
$365\times1024\approx 3.7\times 10^5$ transitions per $\theta$-particle.  Since 
five PMMH steps are performed at each rejuvenation step, the number of calls per $\theta$-particle 
reaches $6\times 10^6$. The dashed line represents a linear regression of
the number of calls against time, indicating that the number of calls seems to
grow linearly in $t$. Note that this linear trend is obtained for a fixed $N_x=1024$.
The quadratic cost of the overall method mentioned in Section~\ref{sec:sub:smc2complexity}
comes from the fact that one would eventually need to increase $N_x$ if observations kept arriving.
Thus if we had two years of daily data instead of one, and if we wanted to obtain the same relative error
in estimating the integrals of Section~\ref{sec:sub:objects},
we would set $N_x$ to $2048$ and the overall expected computational time would be multiplied by four.
Since there are $1024$ $\theta$-particles, the total
number of calls to the functions $f$ and $g$ is in the billions. For the PZ
model, each transition involves solving numerically a differential equation, here
using a Runge-Kutta method RK4(3)5[2R+] as in \cite{murray2013disturbance}. In
wall-clock time, this SMC$^2$ run took about $50$ minutes on a standard desktop
computer with $8$ cores using the optimized software LibBi
\cite{murray2013bayesian}. Across runs, the random occurrence of 
rejuvenation steps incurs random computing times. We collected runtimes between
$40$ and $60$ minutes, using the same algorithmic parameters, over five independent runs.

\begin{figure}
    \centering
    \includegraphics[width=0.9\textwidth]{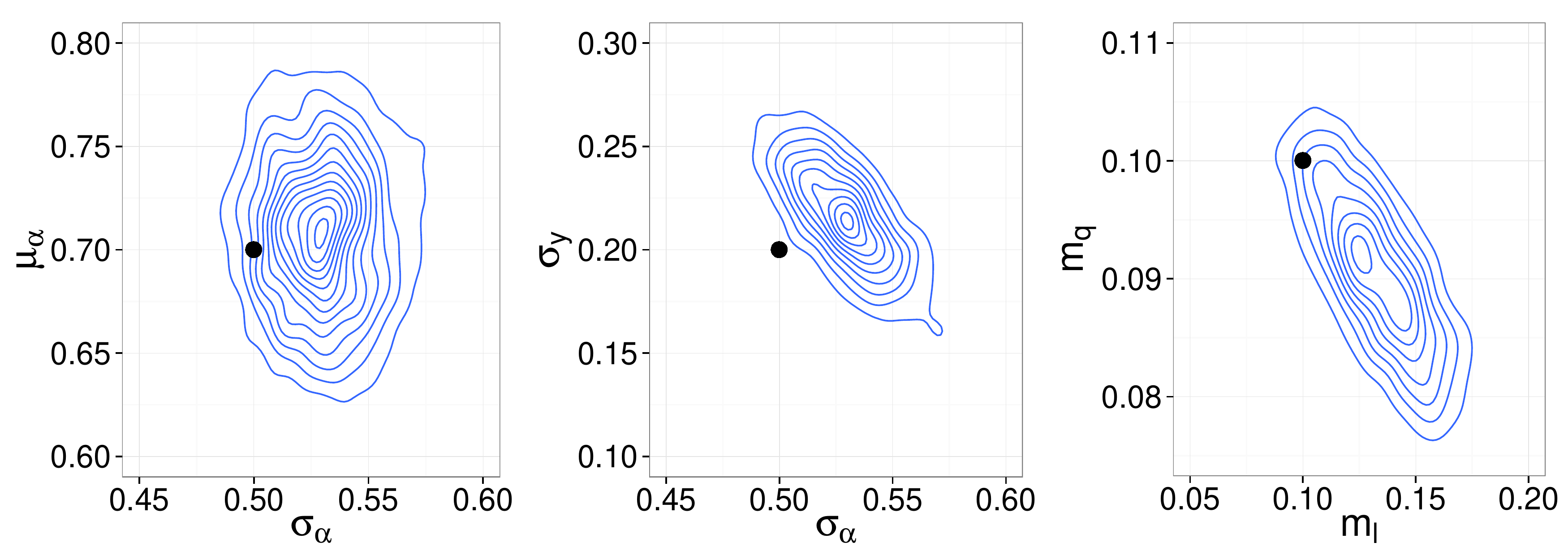}
    \caption{Posterior distribution of the parameters of the PZ model given the
        synthetic dataset of $365$ observations. Contour lines represent the
        estimated density of the pairwise marginal distributions $(\sigma_\alpha,
        \mu_\alpha)$, $(\sigma_\alpha, \sigma_y)$ and $(m_l,m_q)$. The dots indicate
    the values used to generate the dataset.}
    \label{fig:pz:param:density}
\end{figure}

The approximation of the posterior distribution $\pi_{\theta,T}(d\theta)$ at the final time $T=365$
is represented by the pairwise contour plots of Figure~\ref{fig:pz:param:density}.
We see that the posterior distribution concentrates in the neighborhood of the parameter used to generate the dataset, indicated by black dots on the figure.
We note negative correlations between some of the parameters, in particular between $m_l$ and $m_q$, which both explain the instantaneous decrease of the 
zooplankton population size, and between $\sigma_y$ and $\sigma_\alpha$, which both account for the stochasticity of the model.
We observe that the mode of the posterior distribution is not exactly located at the data-generating parameter because the inference
is conditional upon a finite number of observations. Indeed, one could only expect
the data-generating parameter to be recovered when the number of observations goes to infinity. Note also that uniform
priors have been used for all the parameters, therefore the mode of the posterior distribution corresponds exactly to
the maximum likelihood estimate.

\begin{figure}
 \centering
\subfigure[]{\includegraphics[width =  0.4\textwidth]{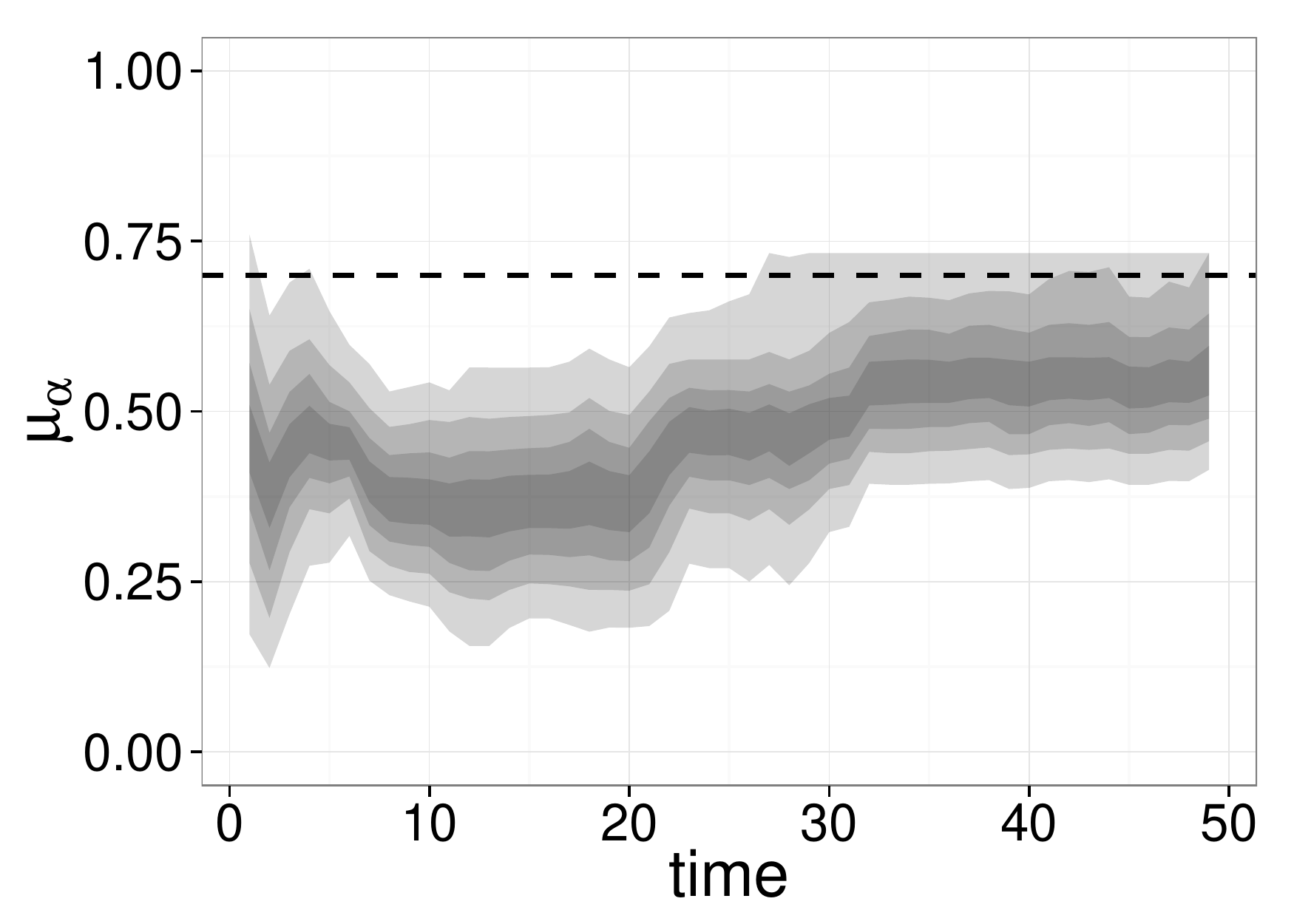}}
 \subfigure[]{\includegraphics[width = 0.4\textwidth]{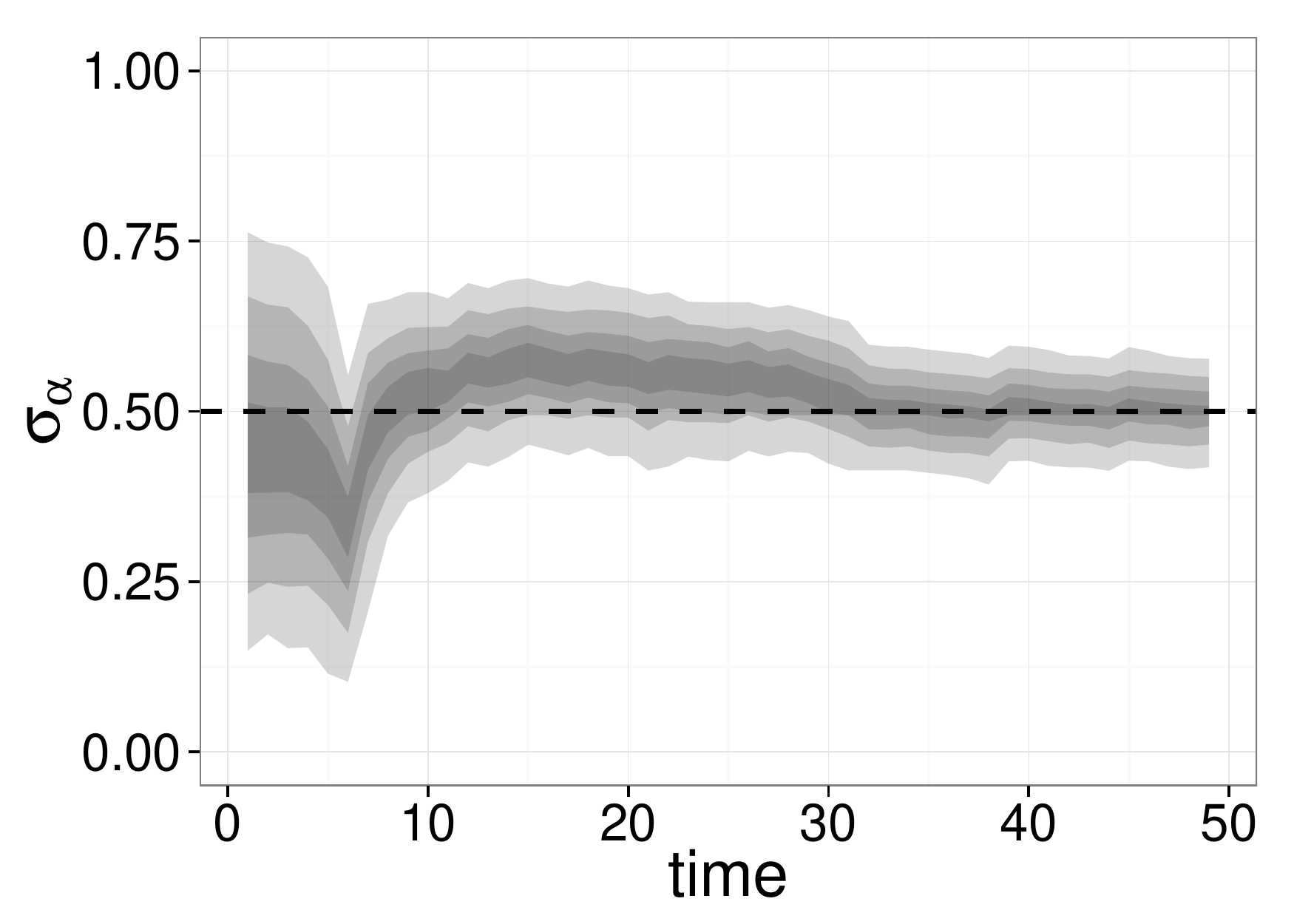}}
 \subfigure[]{\includegraphics[width = 0.4\textwidth]{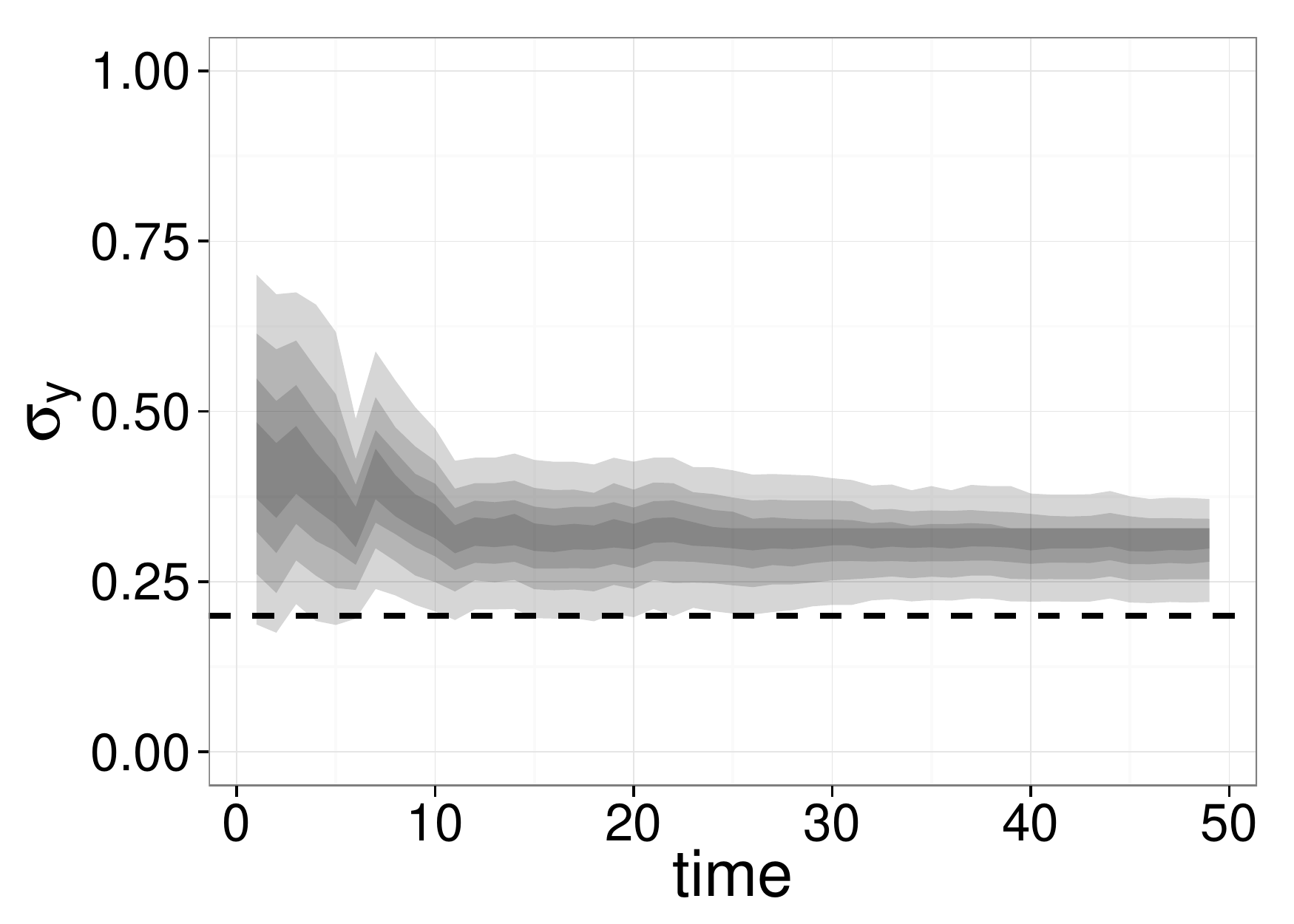}}
 \subfigure[]{\includegraphics[width = 0.4\textwidth]{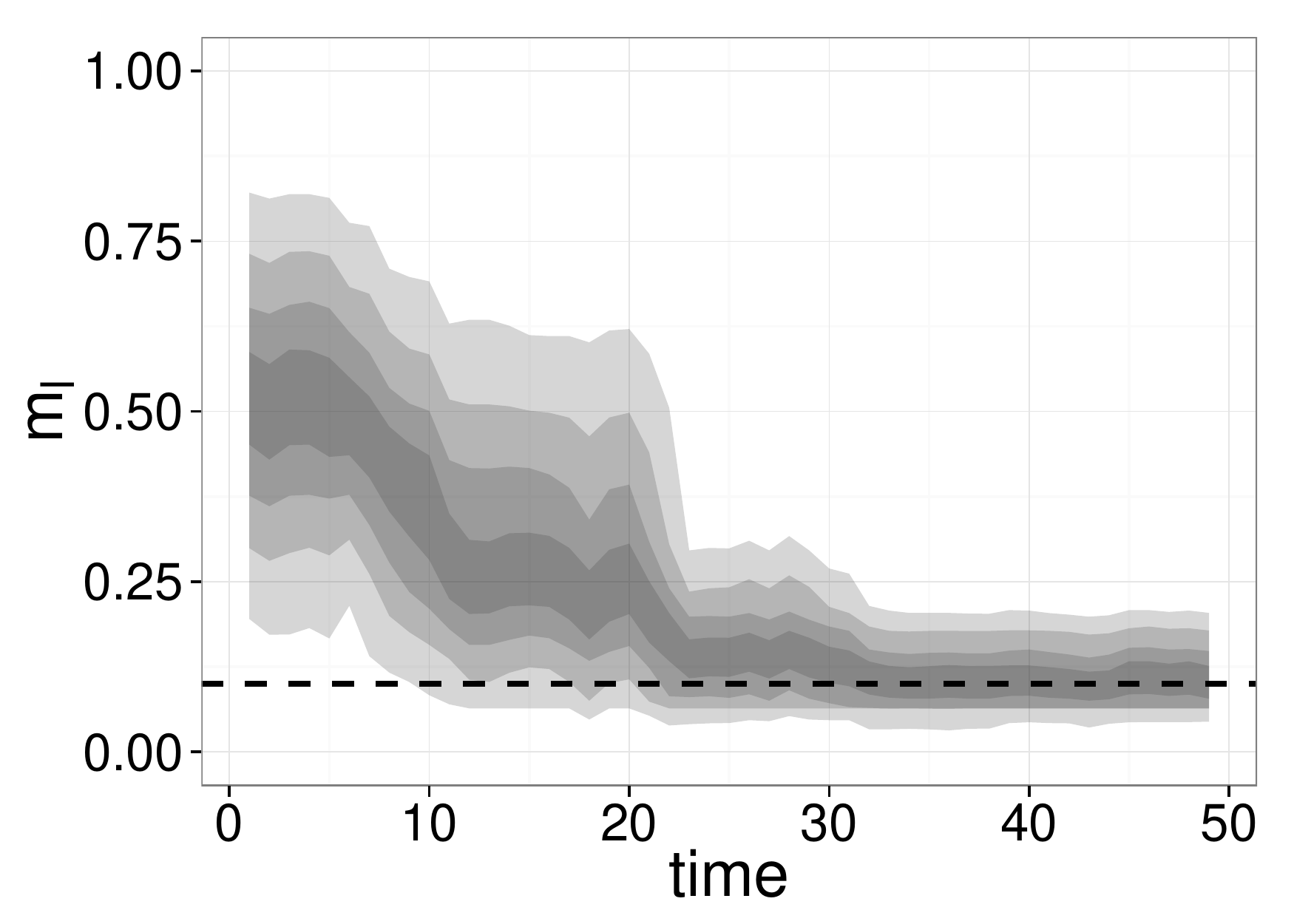}}
 \subfigure[]{\includegraphics[width = 0.4\textwidth]{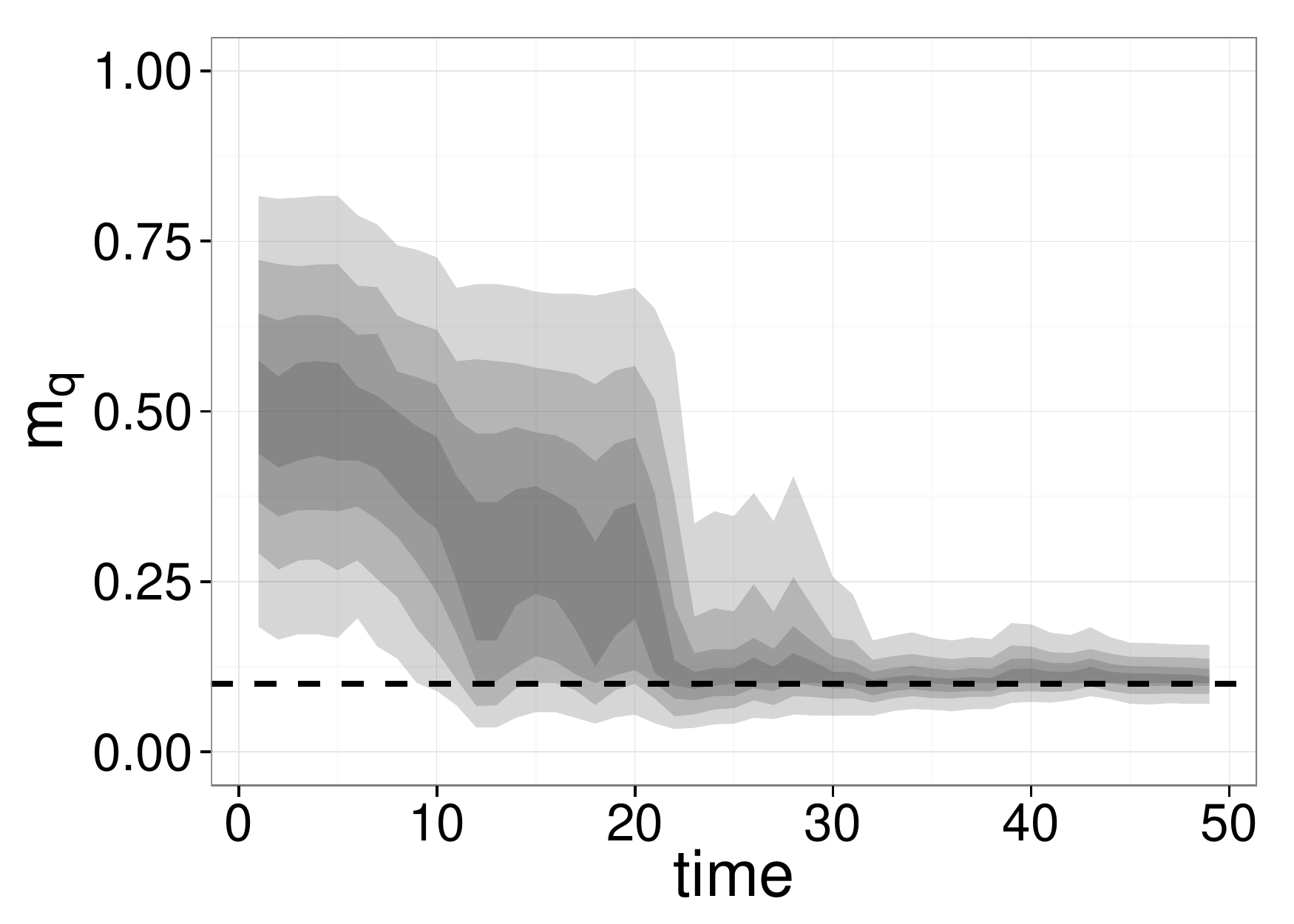}}
    \caption{Evolution of the posterior distribution of each parameter of the PZ model obtained by SMC$^2$, over the first $50$ time steps. The grey ribbons indicate the $10\%,20\%,30\%,40\%,60\%,70\%,80\%,90\%$ quantiles.
    The dashed lines indicate the values used to generate the dataset. }
    \label{fig:pz:param:evolution}
\end{figure}

One advantage of sequential inference is the ability to investigate each intermediate posterior distribution $\pi_{\theta,t}(d\theta)$ for $t = 0, \ldots, T$.
Figure~\ref{fig:pz:param:evolution} represents this evolution for the first $50$ time steps and for each parameter. 
The grey ribbons indicate the $10\%,20\%,30\%,40\%,60\%,70\%,80\%,90\%$ quantiles of each marginal posterior distribution.
The dashed lines indicate the values used to generate the dataset. We see the posterior distributions going nearer the data-generating
parameter as more observations are being assimilated. We also observe that this concentration occurs at a different rate
for each parameter. Indeed, according to asymptotic results on the posterior distribution (Chapter 1 of \cite{ghosh2003springer}), the asymptotic concentration rates
depend on the Fisher information matrix of the model.
In a non-asymptotic setting, as is the case in practice, we could imagine using plots similar to Figure~\ref{fig:pz:param:evolution} to 
guess how many more observations would be needed to reach a given
precision for each parameter.

\begin{figure}
    \centering
    \includegraphics[width=0.9\textwidth]{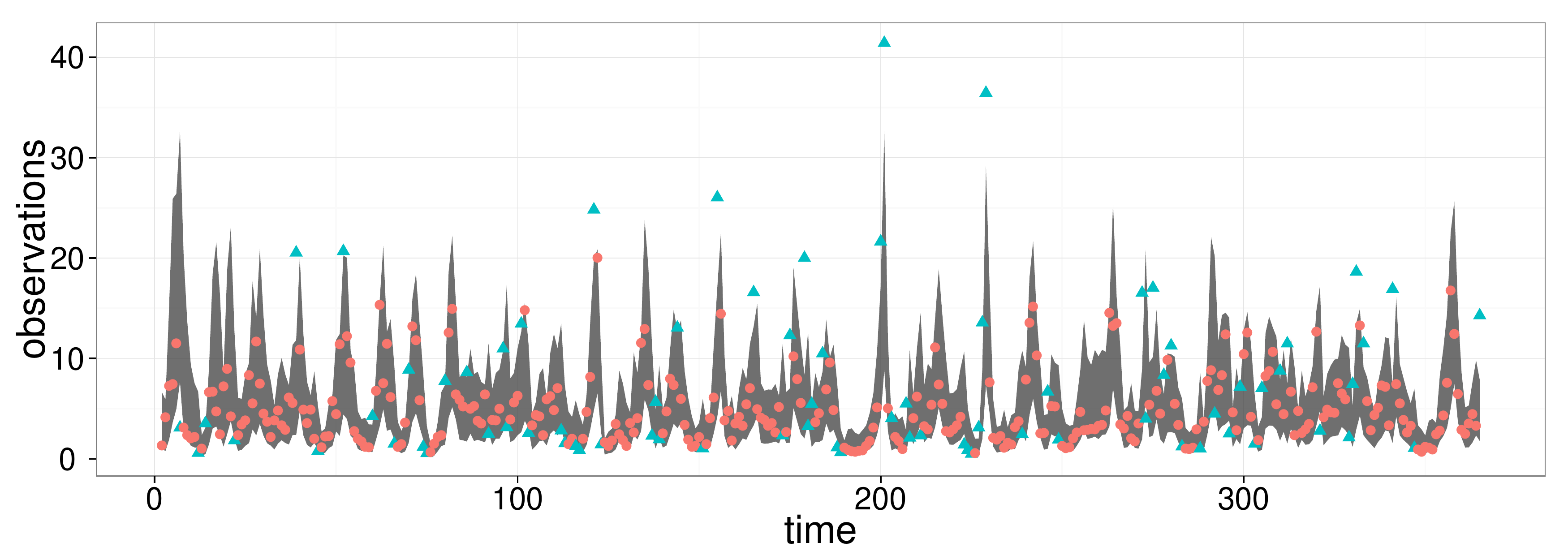}
    \caption{One step predictions obtained using one run of SMC$^2$ on the PZ model. The dark ribbon indicates
        the $80\%$ predictive region of $y_{t+1}$ given $y_{0:t}$ for each time, under parameter uncertainty. 
        Observations that land in the predictive region are indicated by circles, whereas observations landing outside
    are indicated by triangles.}
    \label{fig:pz:predict:obs}
\end{figure}

\begin{figure}
    \centering
    \includegraphics[width=0.9\textwidth]{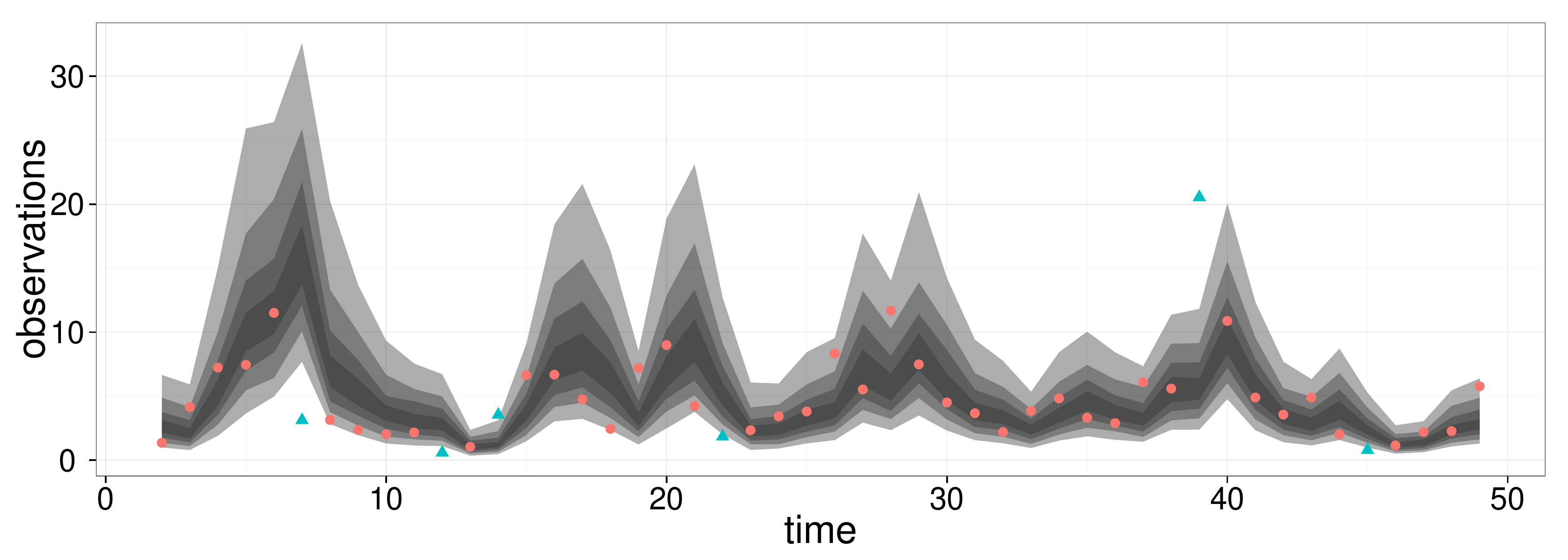}
    \caption{Same as Figure~\ref{fig:pz:predict:obs}, but limited to the first 50 time steps. The grey ribbons indicate the quantiles of one-step predictive regions.}
    \label{fig:pz:predict:obs:zoom}
\end{figure}

The possibility to perform prediction under parameter uncertainty is illustrated on Figure~\ref{fig:pz:predict:obs}. At every time step,
an $80\%$ predictive region is inferred from the particle approximation of $y_{t+1}$ given $y_{0:t}$. The successive regions are joined together
in a grey ribbon. The actual observations are plotted as circles if they fall in the predictive region, and triangles if they fall outside. At the end of the run,
$77$ observations have landed outside the predictive region, which represents $21\%$ instead of the targeted $20\%$. Since the observations are generated from the model,
it is expected that asymptotically in $t$, $20\%$ would fall outside the $80\%$ predictive region.
Figure~\ref{fig:pz:predict:obs:zoom}
is a close-up of Figure~\ref{fig:pz:predict:obs}, focusing on the first $50$ time steps. 

The model evidence estimator of Eq.~\eqref{eq:evidence:smc2estimator}
can be illustrated by introducing another model.  We consider a simplified
model PZ$^\star$, which is defined as model PZ except that the
quadratic mortality term is removed from the 
differential equation:
\begin{align*}
    \frac{dp_t}{dt} &= \alpha p_t - c p_t z_t ,\\
    \frac{dz_t}{dt} &= e c p_t z_t -m_l z_t.
\end{align*}
Thus the parameter is $\theta = (\mu_\alpha, \sigma_\alpha, \sigma_y, m_l)$ and we use the same uniform prior distributions as for the PZ model.
We put a uniform prior over the two models and thus the posterior odds as in Eq.~\eqref{eq:bayesfactor}
reduce to the Bayes factor, $p(y_{0:t}\mid \text{PZ}) / p(y_{0:t}\mid \text{PZ}^\star)$. This ratio can be obtained
by approximating the evidence using the estimator of Eq.~\eqref{eq:evidence:smc2estimator} for each model.
The same algorithmic parameters as described above are used for each model. 

\begin{figure}
    \centering
    \includegraphics[width=0.9\textwidth]{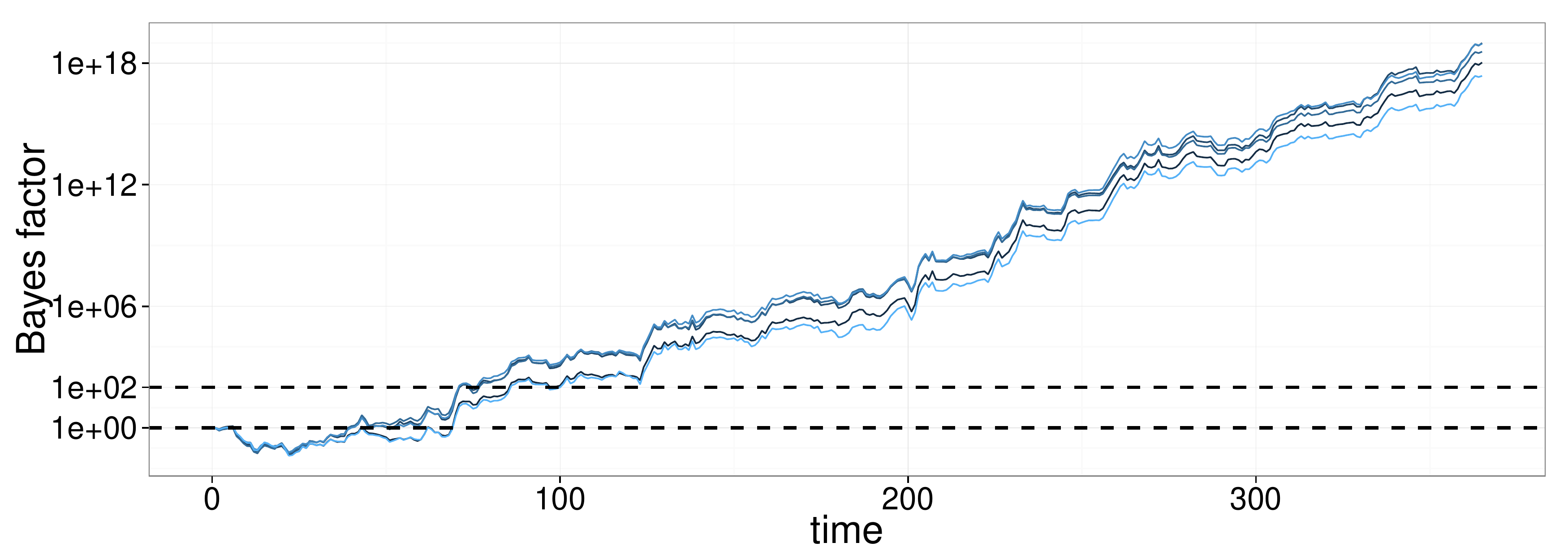}
    \caption{Bayes factor of the PZ model versus the simplified PZ$^\star$ model against time. The bottom dashed line
    indicates $1$ and the top one indicates $100$. Values larger than $1$ indicate support for the PZ model. The full lines correspond to the estimates of the Bayes factor for five independent runs
    of the SMC$^2$ algorithm.}
    \label{fig:pz:bayesfactor}
\end{figure}

\begin{figure}
    \centering
    \includegraphics[width=0.9\textwidth]{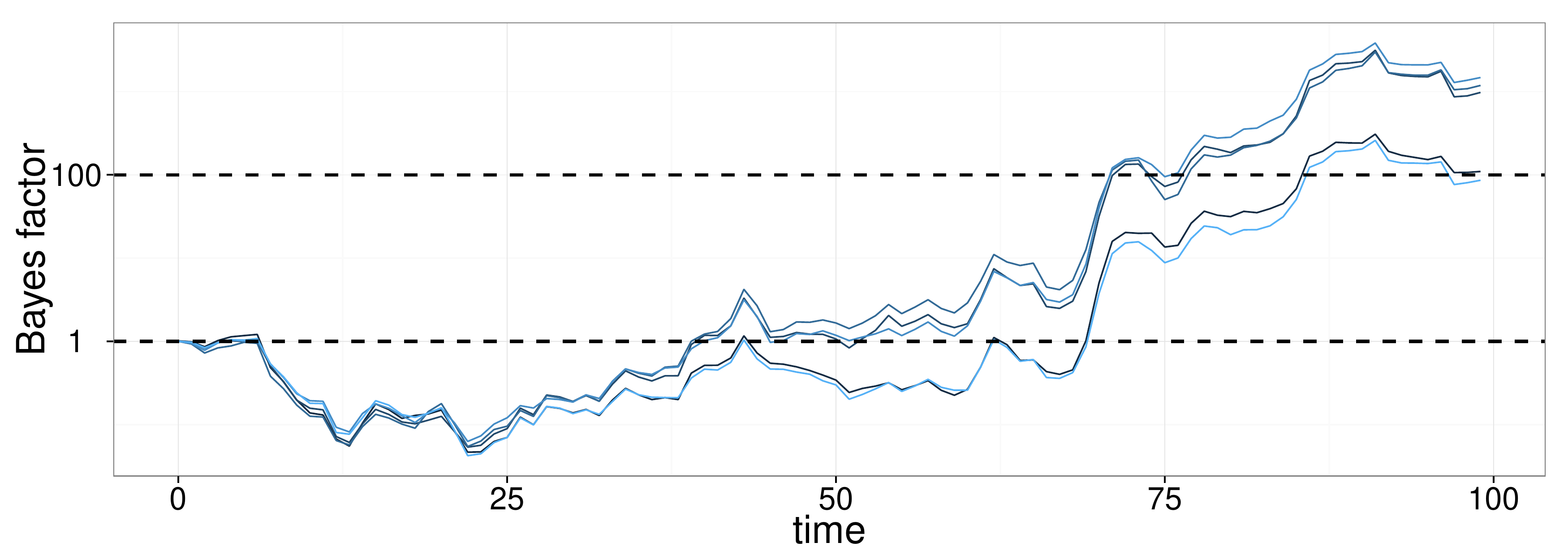}
    \caption{Same as Figure~\ref{fig:pz:bayesfactor}, but limited to the first
    $100$ time steps. Initially the simpler PZ$^\star$ model is preferred by the
    Bayes factor, but after $100$ observations the more complex, data-generating PZ model becomes
    strongly supported by the criterion.}
    \label{fig:pz:bayesfactor:zoom}
\end{figure}

Figure~\ref{fig:pz:bayesfactor} shows the Bayes factors against time, obtained
from five independent runs, and Figure~\ref{fig:pz:bayesfactor:zoom}
is a close-up on the first $100$ time steps. The bottom horizontal dashed line indicates
$1$. A Bayes factor of $1$ indicates no support of the data for one model
compared to the other. Values close to zero support model PZ$^\star$
while values larger than one support model PZ.
The graph shows that for the first $50$
observations, model PZ$^\star$ seems supported by the data. However with
more observations, the Bayes factor starts supporting the PZ model,
and after time $100$ each of the five independent runs estimates the factor above
$100$. The factor keeps increasing to extremely large values as more observations
are assimilated. Here the dataset is generated using the PZ model so the end
result does not come as a surprise. The sequential Bayes factor estimation 
shows Occam's razor principle in action, as mentioned in Section
\ref{sec:sub:objects}: the simpler model is favoured when few observations are
available. Here about $100$ data points are enough to choose the data-generating model with confidence,
according to the Bayes factor criterion.
Note how the five independent evidence estimates diverge from each other as observations accrue, showing that the estimator is not stable over time for a
fixed number of particles $N_x$ and $N_\theta$. This confirms that the evidence estimator from SMC$^2$ is not online.

\section{Discussion\label{sec:discussion}}

Let us discuss again the objects of inference of Section~\ref{sec:sub:objects} in the light of
the reviewed methodology.  For a given parameter value, filtering and prediction as in the integrals of Eq.~\eqref{eq:filtering},~\eqref{eq:pred:states},~\eqref{eq:pred:obs} can be
approximated in an online manner using particle filters, as described in
Section~\ref{sec:methods}.  
There are currently no online and exact methods, in the sense of  
Section~\ref{sec:sub:blackbox}, that take into account parameter uncertainty 
as in Eq.~\eqref{eq:posterior},~\eqref{eq:filtering:uncertainty},~\eqref{eq:evidence}.
The SMC$^2$ method
proposed in \cite{chopin2013smc2} and \cite{fulop2013efficient}, and described
in Section~\ref{sec:smcsamplers}, is sequential, in the sense that the
estimators can be updated upon the arrival of new observations. However the incremental cost of the algorithm has to
grow linearly with $t$ in order to control the relative variance of the estimators. Hence a complete run of the algorithm has a quadratic
cost in the length of the time series, and thus is not applicable for long time
series. Informally, for the PZ model used as an illustration in Section~\ref{sec:numerics}, 
the SMC$^2$ algorithm runs in a reasonable time on standard hardware, for thousands of observations, but not for millions. 
The numbers would change with the model and the application,
but in general online inference under parameter uncertainty is still an open question and an active area of research; see \cite{crisan2013nested,zhou2015biased}
for recent developments.

One of the difficulties comes from the likelihood estimator of Eq.~\eqref{eq:likelihood:estimator},
which requires a quadratic cost in the number of observations to guarantee a bounded relative error. On the other hand, the Kalman filter yields likelihood evaluations in a linear cost, 
but only for linear Gaussian models. It is unclear whether an intermediate setting exists, where the likelihood could be estimated in a super-linear but sub-quadratic cost, at least for some models.

For time series of moderate length, filtering, prediction and parameter inference are still 
challenging problems when the dimension $d_x$ of the state space
$\X$ is large. Indeed the variance of standard particle filter estimators typically increases exponentially with $d_x$. Recent developments such as \cite{rebeschini2013can,beskos2014stable}
could help scaling particle methods to larger dimensions, with many potential applications 
in spatial state space models \cite{reich2015}.
Another issue specific to large-dimensional state space models is the large computer memory
required to store the particles, and especially the paths ($\bar{x}_{0:t}^{1:N_x}$ in the notation of Section~\ref{sec:sub:methods:filtering}). Large memory usage
also involves large communication costs on distributed hardware, whenever particles have to be
sent from one machine to another.
The expected memory usage of storing the paths has been studied in \cite{jacob2013path}.
Methods to reduce the memory and communication costs on distributed hardware have been proposed in \cite{bouchardcote2012,bouchardcote2014}.

Another challenge is to adapt computational methods to larger classes of models. The plug and play methods described in Section~\ref{sec:methods} are compatible with parametric models, where the latent process can be simulated and the measurement density can be evaluated point-wise. On top of hidden Markov models, particle filters can be implemented for non-Markovian models, as long as these two requirements are met. Inference in non-Markovian models using particle methods has been recently considered in \cite{lindsten2012ancestor}. The performance of particle methods in non-Markovian settings has been partially studied in \cite{chopin2011stability}.
Recent applications of non-Markovian models, for instance in probabilistic programming \cite{wood2014new}, motivate further research in this direction.

A number of articles have considered non-parametric hidden Markov models.
The authors of \cite{caron2008bayesian} consider linear models with a transition equation of the form
$x_{t} = A_t x_{t-1} + G_t v_t$, where the distribution of the noise $v_t$ is modelled with
a Dirichlet process mixture. A non-parametric model is also considered for the measurement noise.
Estimation is then performed using Markov chain Monte Carlo or sequential Monte Carlo methods.
In the more recent literature, \cite{frigola2013bayesian} consider a transition equation of the form $x_{t} = f(x_{t-1}) + v_t$ 
and put a Gaussian process prior on the function $f$; particle Markov chain Monte Carlo
methods then enable inference under parameter uncertainty.
Note that combining non-parametric models for the function $f$ and for the noise $v_t$ is not obvious
because of identifiability issues.
Other instances of non-parametric hidden Markov models
consider the case where the hidden process lives on an infinite but discrete state space
\cite{teh2006hierarchical,fox2008hdp}. Particle Markov chain Monte Carlo methods have recently been used in this context \cite{tripuraneni2015linear}.
The case of finite state space and non-parametric specification of the measurement distribution is considered in \cite{yau2011}. 
Sequential algorithms to perform inference in continuous space, non-linear, non-parametric hidden Markov models would constitute an interesting addition to the current methodology.

\begin{acknowledgement}
The author gratefully acknowledges EPSRC for funding this research through
grant EP/K009362/1, thanks the organizers of the Journ\'ees MAS 2014, thanks Arnaud Doucet, Lawrence Murray and Aimee Taylor for useful
comments. This article is dedicated to the memory of Philip Perry and some long
discussions on the Bayesian approach. 
\end{acknowledgement}

%%-----------------------------
%%      your bibliography
%%-----------------------------
\bibliographystyle{plainnat}
\bibliography{complete}
\end{document}